\documentclass[a4paper,11pt]{article}
\pdfoutput=1 

\usepackage{jheppub} 

\usepackage[T1]{fontenc} 
\usepackage[utf8]{inputenc}
\usepackage{amsfonts}
\usepackage{graphics}
\usepackage{epsfig}
\usepackage{graphicx,wrapfig,float}
\usepackage{xcolor}
\usepackage{multirow}
\usepackage{lineno}
\newcommand {\be}{\begin{equation}}
\newcommand {\ee}{\end{equation}}
\newcommand {\ba}{\begin{eqnarray}}
\newcommand {\ea}{\end{eqnarray}}
\newcommand {\invfb}{$fb^{-1}$}
\newcommand {\tanb}{$\tan\beta~$}
\newcommand {\ra}{\rightarrow}

\title{\boldmath Observability of $s$-channel Heavy Charged Higgs at LHC Using Top Tagging Technique}

\author[]{Majid Hashemi,}
\author[]{Gholamhossein Haghighat}

\affiliation[]{Physics Department,\\College of Sciences, Shiraz University, Shiraz 71454, Iran}

\emailAdd{hashemi$\_$mj$@$shirazu.ac.ir}
\emailAdd{hosseinhaqiqat@gmail.com}
\abstract{
In this paper, the question of observability of a heavy charged Higgs in the mass range 400 GeV $<m_{H^{\pm}}<$ 1000 GeV, is addressed. The production process is set to $pp \ra H^{\pm} \ra t\bar{b}$ at 14 TeV LHC. The analysis benefits from top tagging technique which is based on finding a fat jet as a result of the boosted top quark decay in signal events. A detailed hadron level analysis is performed and selection efficiencies are presented with different charged Higgs mass hypotheses. Finally running toy experiments and using pseudo-data, a fit over signal plus background distributions is performed to assess possibility of reconstructing the charged Higgs peak and its invariant mass measurement. It is shown that the charged Higgs mass can well be reconstructed in the mass range 500 GeV to 1 TeV, with a signal significance which depends on \tanb. Eventually 5$\sigma$ discovery and 95$\%$ C.L. exclusion contours are also provided. }

\keywords{Beyond Standard Model, Supersymmetry, Charged Higgs}
\begin{document}
\maketitle
\flushbottom

\section{Introduction}
The charged Higgs boson is one of the particles whose observation would reveal the existence of models beyond the Standard Model of Particle Physics. Within the framework of Minimal Supersymmetric Standard Model (MSSM), the Large Hadron Collider (LHC) is currently performing search for this particle in the parameter space ($m_{H^{\pm}},$\tanb) with no evidence still found. Here, \tanb is the ratio of vacuum expectation values of the two Higgs doublets used to make the MSSM \cite{hhg}.

The theoretical and phenomenological studies which paved the way to experimental searches include studies of properties of the charged Higgs \cite{CHOld}, observability at Tevatron \cite{CHDPRoyFermilab}, $\tau$ polarization studies \cite{taupol,CHRoy1,CHRoy2,CHRoy3,CHSharpening}, study of the $gb$ fusion as a source of charged Higgs production \cite{gbCH}, the $tb$ decay channel study \cite{tbSuperCollider}, heavy charged Higgs study at LHC \cite{HeavyCHatLHC},  charged Higgs three body decay \cite{CH3body}, pair production in gluon-gluon collisions \cite{CHPair}, triple $b$-tagging as a tool for charged Higgs detection \cite{CH3b}, charged Higgs production in $bg$ fusion \cite{bgCH2,gbth2005} and the associated production of top and charged Higgs \cite{tH}. These studies, all together, showed that the promising decay channel for a heavy charged Higgs is $t\bar{b}$ and $\tau\nu$, while for a light charged Higgs (lighter than the top quark), decay to $\tau\nu$ plays the main role. Of course when a $\tau$ lepton is produced in the event, use of the $\tau$ polarization makes it easy to distinguish between the charged Higgs boson signal and SM background events.

The more data LHC collects, the higher charged Higgs masses are accessible for analysis. This is due to the fact that all production processes are decreasing functions of the charged Higgs mass. Therefore charged Higgs masses near the TeV scale may need a large amount of data to observe and assuming its existence, the heavier the charged Higgs, the more challenging its observation. This is the reason of attempt, in this paper and few upcoming ones, for analysis of charged Higgs near the TeV scale. The result of this analysis is expected to help LHC experiments in their progress to this region of parameter space. LHC data will soon provide opportunity to explore heavy charged Higgs boson and the analysis to be presented in this paper and similar ones would reveal their importance when working in that area of parameter space. It may be notorious that accroding to flavour Physics studies and global fits to experimental data, a charged Higgs with $m_{H^{\pm}}\simeq 600$ GeV at low \tanb values fits well with EW and flavour Physics observations and the overall deviation of theory (2HDM type II) from experiment in terms of a global ${\chi}^2$ is less than that of SM \cite{mahmoudi}.

In order to establish where we are and where we are going, a presentation of current results of collider searches for the charged Higgs follows. Currently there are several main limits on the mass of the charged Higgs. There is a \tanb independent low mass limit of $m_{H^{\pm}}>78.6$ GeV from direct searches at LEP \cite{lep1,lep2,lepexclusion1} while their indirect searches set a limit of $m_{H^{\pm}}>125~ \textnormal{GeV}$ \cite{lepexclusion2}. The latter is used in this analysis as it covers the former. The Tevatron searches by D0 \cite{d01,d02,d03,d04} and CDF Collaborations \cite{cdf1,cdf2,cdf3} exclude high \tanb values. The scope of such analyses is the light charged Higgs which is now confirmed and extended by LHC searches.

LHC experiments (CMS and ATLAS) started light and heavy charged Higgs analyses based on Monte Carlo simulations in separated works. Light charged Higgs possibility of observation was studied by CMS \cite{MyLightCH} and ATLAS \cite{BiscaratLightCH} using the top pair production process with a top quark decaying to charged Higgs. Such an assumption is, however, almost excluded according to LHC real data analyses performed by CMS \cite{LightCH7TeVCMS1,LightCH7TeVCMS2} and ATLAS  \cite{LightCH7TeVATLAS1,LightCH7TeVATLAS2}. Their conclusion is that a charged Higgs with $m_{H^{\pm}}<160$ GeV is excluded at 95$\%$ C.L. for almost all \tanb values.

Heavy charged Higgs Monte Carlo analyses at CMS \cite{SLowetteHeavyCH,RKHtaunu} and ATLAS \cite{KAHeavyCH1,KAHeavyCH2} have been based on associated production of the charged Higgs and a top quark, i.e., $g\bar{b} \ra t H^{-}$, with special care of both $H^{\pm}\ra \tau\nu$ and $H^{\pm}\ra t\bar{b}$ decay channels. The ongoing LHC direct searches also rely on the same production process and decay in the search for heavy charged Higgs \cite{CHtaunu8TeVCMS,CHtaunu8TeVATLAS}. There has been also indirect searches for the charged Higgs through the observation of deviation from what is expected from SM $t\bar{t}$ events \cite{CHIndirect8TeVATLAS}. However result of the analysis presented in \cite{CHIndirect8TeVATLAS} is superceded by direct search results \cite{CHtaunu8TeVCMS,CHtaunu8TeVATLAS} which exclude heavy charged Higgs masses in the range $m_{H^{\pm}}=180-230$ GeV at high \tanb values.\\

\section{Single Top Events as Sources of Charged Higgs}
In a couple of previous works, single top events were proved to be viable sources of light and heavy charged Higgs in $t$-channel and $s$-channel single top production respectively \cite{st1,st2}. The heavy charged Higgs can be produced as a resonance in $s$-channel single top production, i.e., $pp \ra H^{\pm} \ra t\bar{b}$ with the possibility of reconstructing the charged Higgs invariant (transverse) mass in hadronic (leptonic) final state. The analysis presented in \cite{st2} relies on the leptonic final state which is related to the leptonic decay of the W boson in the top quark decay.  It has led to promising results relevant to $200 ~\textnormal{GeV}<m_{H^{\pm}}<400~ \textnormal{GeV}$. However, the analysis fails for higher charged Higgs masses at low luminosity of LHC due to the small cross section of production process. Including off-diagonal contribution of incoming quark pair has also been shown to improve the results sizably \cite{myplb}. However, one may need to use technical tools for heavier charged Higgs bosons in order to improve the signal significance and extend the analysis to TeV scale charged Higgs.

In what follows, we try to describe the top tagging technique and its application to selection of signal events which contain a boosted top quark. The top quark may get a large boost in decay of a heavy particle from beyond Standard Model, like MSSM charged Higgs, which decays to $t\bar{b}$ if its mass is above the kinematic threshold. As an example, a top quark from a charged Higgs decay with $m_{H^{\pm}}=700$ GeV gets a much larger Lorentz boost compared to what it acquires in SM events like $t\bar{t}$. Identifying such top quarks using top tagging technique which is based on sub-jet identification in the top quark hadronic decay may help a lot in discrimination of BSM events and their characteristic particles.

\section{Jet substructure and top tagging technique}
Due to its high center of mass energy, LHC produces highly boosted particles in its detectors frames. When such energetic particles decay, their decay products tend to be collinear due to the Lorentz boost they receive at the laboratory rest frame. In case of top quark which decays before it can hadronize \cite{topdecay}, this situation results in collinear jets in the fully hadronic decay $t\ra W^{+}b \ra jjb$, i.e. , two light jets from the W boson decay and a $b$-jet directly from the top quark decay. Identification of such jets by present jet reconstruction algorithms is a challenging task due to not being well separated from each other. A cone of $\Delta R=\sqrt{(\Delta \eta)^2+(\Delta \phi)^2}=0.4$ for jet reconstruction can not distinguish well each one of the three jets from the top quark decay. In this case, a top quark decay appears as a ``fat'' jet spreaded in a large cone of $\Delta R\simeq 1.5$. Therefore a jet algorithm should start from a large cone of that size, identify the fat jet, and preferably check the invariant mass of ``sub-jet'' pairs. By definition, a sub-jet is a jet inside the large cone of the highly boosted particle, in this case the top quark, associated with either the $b$-jet or the two light jets from the W hadronic decay.

From the Physics point of view, few percent of top quarks from Standard Model (SM) events may acquire enough boost to appear as a fat jet \cite{JohnHopkins}. However, a massive $s$-channel resonance from beyond SM, may produce highly boosted top quark in its decay. An example of such a case, relevant to this work, is a massive $s$-channel charged Higgs production, $pp \ra H^{\pm} \ra t\bar{b}$ whose final state particles are similar to the $s$-channel SM single top but with a large difference in kinematics. A charged Higgs with a mass of a reasonable fraction of TeV, produces a top quark which is well distinguishable from the $s$-channel SM single top. Such a jet identification and reconstruction receives a negligible fake rate from QCD multi-jet events and electroweak gauge boson production processes due to the high kinematic threshold used for BSM signal events.

In recent years, special care has been devoted to the so called ``top tagging'' technique. With the more data LHC provides to its experiments, CMS and ATLAS, searches for heavier BSM particles becomes feasible. In some cases, searching for a heavy BSM particle requires not only more data but also special care in experimental analysis due to the high particle multiplicity and the associated combinatorial background. The high particle multiplicity may be the result of heavy electroweak gauge boson decays as well as top quark decays in their hadronic final state. The case of heavy charged Higgs, stated previously, is an example of such a situation. The top tagging algorithm has been designed to identify top quarks in such highly boosted regimes. It tries to deal with not only the high particle multiplicity in events, but also the collinearity of final state jets from the top quark decay. One of the early reports on introducing and applying boosted jet tagging technique is the one known as BDRS tagger \cite{BDRS}. It was originally designed for improving searches for associated Higgs production in $t\bar{t}H,~WH$ and $ZH$ for the low mass Higgs when the Higgs boson decays to a $b\bar{b}$ pair. It was a successful application of tagging technique in which a large cone for the jet reconstruction was used to identify a fat jet and decompose it to two $b$-jets. This technique was shown to work better than the standard reconstruction techniques, due to moving the analysis to a highly boosted regime. Although only about $5\%$ of events are produced with a reasonable boost (Higgs boson $p_T~>~200$ GeV), several advantages led to a more promising result in that analysis: the larger $VH$ system mass resulted in more central events with decay products having higher transverse momenta to be tagged. SM background events were less likely to produce a high-$p_T$ $b\bar{b}$ system consistent with the Higgs boson mass. Working in the highly boosted regime provided the opportunity to make the $ZH$ process with $Z\ra \nu\bar{\nu}$ visible because of the large missing transverse energy.

In the current analysis, the same physics reasons motivate using a top tagger. A standard search may be dominated by the large QCD events due to their large cross section. Applying a high $p_T$ threshold for jets, reduces this background dramatically. There are also irreducible backgrounds from SM single top events (especially the $s$-channel single top) which are suppressed only at a highly boosted regime in which it is almost impossible for the virtual W boson to produce a $t\bar{b}$ pair with a high-$p_T$ top quark. Using a top tagger, the jet combination is also performed better for final state reconstruction. In the current analysis, there are two $b$-jets in the event. The top tagger tags the $b$-jet inside the fat jet cone as the one from the top quark decay. Using this information, one can reconstruct the top quark invariant mass using jets inside the top tagging cone and reserve the second $b$-jet for the final charged Higgs reconstruction. In a standard search, such combinatorial issues may reduce the signal to background ratio.

There are a large number of articles related to the concept of top tagging from the point of view of the algorithm development and its application to signal selection at LHC. The idea of identifying jet substructure has been discussed in \cite{ThalerWang,jss2,SeatleTagger1,SeatleTagger2} while its application in Higgs boson searches (SM and MSSM) can be found in \cite{BDRS,jss6,jss8}.

The general idea of jet substructure is based on finding at least three sub-jets in the hard jet cone satisfying a $p_T$ threshold criterion. The starting point is clusterizing input particles into hard jets with a jet algorithm in a large cone, typically, of the size $\Delta R = 1.5$. The hard jet is a candidate for the top quark. Therefore it is required to pass a kinematic cut defined as $p_T>200$ GeV and $\eta<2.5$ where $\eta=-\ln\tan(\theta/2)$ and $\theta$ is the polar angle with respect to the beam axis. For moderately boosted tops from, e.g., a charged Higgs in the mass range $500~ \textnormal{GeV}<m_{H^{\pm}}<1000~\textnormal{GeV}$, decaying to the top quark, the $p_T$ cut can be lowered to 150 GeV \cite{jss8}. The clustering sequence for the jets is then used for decomposition in a primary step where two parent clusters are identified as candidates for the jet pair from W boson decay and the $b$-jet from the top quark decay. The secondary decomposition then tries to find grandparent clusters by decomposing one of the parent clusters into two, resulting in a final number of  at least three clusters. These final clusters are considered as sub-jets from the top quark decay and their invariant mass is required to satisfy the top quark mass window. The invariant mass of the three sub-jets is referred to as the hard jet mass. There are also requirements applied on the jet pairs to check if the invariant mass of one pair satisfies the W boson mass window as well as a threshold requirement applied on the invariant mass of the jet pair with minimum invariant mass. Details of the cuts, though being algorithm dependent, can be found in \cite{JohnHopkins}.

In what was said above, the jet substructure and its application to top tagging was described briefly. However, one may also refer to \cite{YSplitter} for a detailed description of the algorithm. The application of fat jets to Higgs searches has been addressed in \cite{tt3}. The stop reconstruction using top tagging technique has been reported in \cite{HEPTopTagger1}. Tagging single tops has also been proved to be promising using top tagging algorithm in \cite{tt5}. There are also reports on how to update the top tagging algorithm for a better fake rate reduction \cite{HEPTopTagger2,tt7}.

There are several algorithms already available for top tagging. The BDRS \cite{BDRS} is one of the early algorithms based on mass drop criterion in the jet un-clustering. It is originally designed for Higgs decay to bottom quark pair and uses C/A jet algorithm. The YSplitter algorithm \cite{YSplitter} is based on the structure of splitting history and is originally developed by ATLAS collaboration for $Z'$ searches and is useful for heavy particles with $p_T~>~300$ GeV. Contrary to the case of YSplitter, the Seatle Tagger \cite{SeatleTagger1,SeatleTagger2} is based on removing all soft and collinear splittings of QCD jets and studying the remaining massive splittings (the so called "pruning" procedure). The John Hopkins Tagger \cite{JohnHopkins} is one of the first tagging algorithms useful for a two step top decay, starting with a top quark transverse momentum threshold of 1 TeV. It is useful for highly boosted top quarks. There is also Thaler-Wang Tagger \cite{ThalerWang} based on jet energy drop and is useful for top $p_T~>~800$ GeV.

The HEPTopTagger \cite{HEPTopTagger1,HEPTopTagger2} is a development of BDRS method for multi-step top decays. It is originally desigend for $t\bar{t}H$ searches but can be used for other processes. It starts from a considerably lower top $p_T$ and is thus useful for moderately boosted top quarks. While BDRS method relies on the symmetry of the heavy particle decay, the HEPTopTagger does not require a symmetry in the top quark decay. According to the above introduction and comparison of algorithms, the HEPTopTagger is used in this analysis as the top tagging algorithm.

\begin{figure}
 \centering \includegraphics[width=0.7\textwidth]{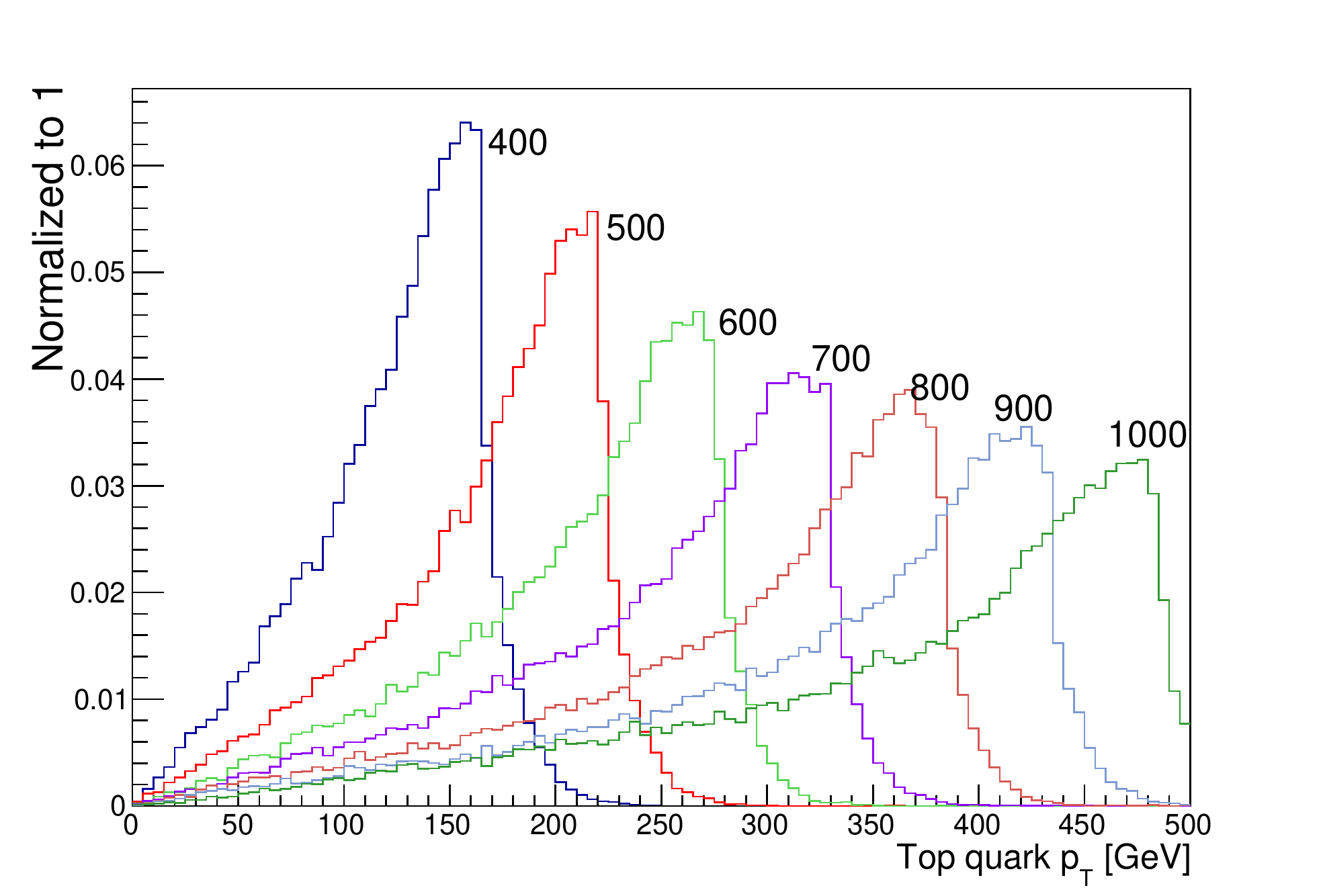}
 \caption{The top quark transverse momentum in signal events for different charged Higgs mass hypotheses. Numbers on histograms correspond to the charged Higgs mass in GeV. \label{toppt}}
 \end{figure}

The tagger is said to be useful for top quark $p_T$ down to 200 GeV \cite{tt1}. However, as will be shown, a reasonable tagging efficiency and background suppression is achieved for top quarks with even less $p_T$. In the current analysis, two thresholds of 150 and 200 GeV for the top quark $p_T$ were tested leading to the conclusion that the cut at 150 GeV is more suitable for this analysis. There are several reasons for that. The cut at 150 GeV avoids the signal statistics reduction especially for low charged Higgs masses for which the cross section is small. Applying $p_T~>~200$ GeV reduces both signal and background without improving the signal significance. On the other hand it is not useful for charged Higgs masses as low as 400 GeV for which the top quark $p_T$ is below 200 GeV and applying a cut at 200 GeV suppresses almost all the signal. Figure \ref{toppt} shows the top quark $p_T$ in signal events with different charged Higgs mass hypotheses. The top tagging has a low efficiency for top quarks produced from $m_{H^{\pm}}=400$ GeV, however, it is still able to identify the top quark and its decay constituents, if a threshold of 150 GeV is applied. Therefore, $p_T~>~150$ GeV was adopted for the top quark transverse momentum in this analysis.
\section{Signal and Background Events and their Cross Sections}

\begin{figure}
 \centering \includegraphics[width=0.7\textwidth]{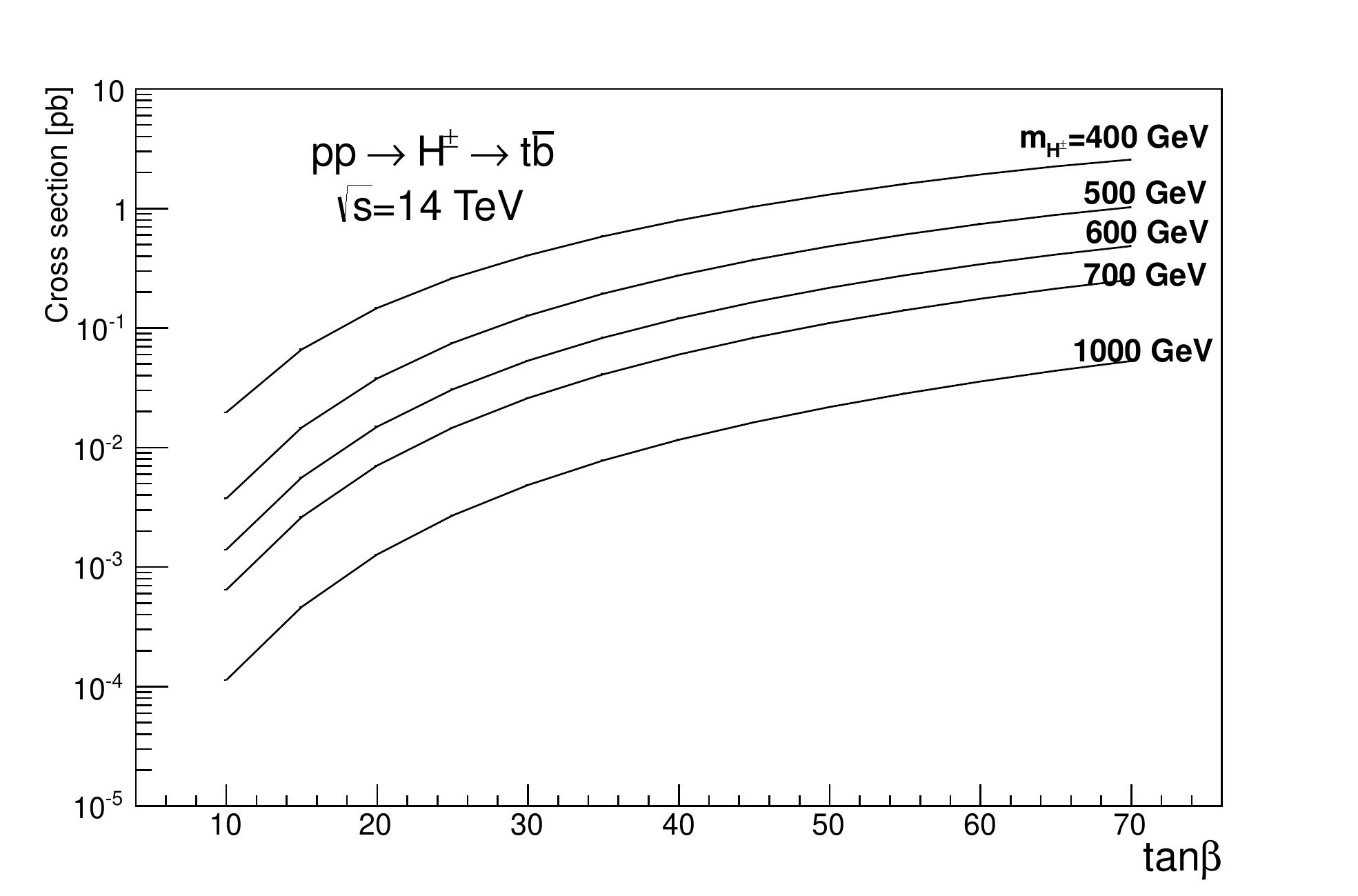}
 \caption{The signal production cross section as a function of $\tan\beta$. \label{Xsecs}}
 \end{figure}

The signal is defined as a process of single charged Higgs production in $s$-channel, followed by the charged Higgs decay to a pair of top and bottom quarks. The theoretical framework is MSSM, $m_{h}$-max scenario \cite{lepexclusion2} with the following parameters: $M_{2}=200$ GeV, $M_{\tilde{g}}=800$ GeV, $\mu=200$ GeV and $M_{SUSY}=1$ TeV. The signal cross section depends on the charged Higgs mass as well as \tanb as shown in Fig. \ref{Xsecs}. These values are obtained using CompHEP 4.5.2 \cite{comphep1,comphep2} with charged Higgs widths taken from FeynHiggs 2.8.3 \cite{feynhiggs1,feynhiggs2,feynhiggs3}. Quark masses are taken from Particle Data Group \cite{pdg}. What is observed here is that the charged Higgs production cross section decreases with increasing its mass, however the reduction can partially be compensated by increasing \tanb. On the other hand, heavy charged Higgs bosons, may lead to invariant mass distributions lying on the tail of the background distribution thus compensating again partially the small signal cross section in that region. Therefore the two factors of signal cross section and the invariant mass distribution play competing roles.

The main background processes are top quark pair production, single top ($s$-channel and $t$-channel production) and single gauge boson W+jets production. The cross section of the top production processes are calculated at NLO using MCFM 6.1 \cite{MCFM1,MCFM2,MCFM3,MCFM4}, while PYTHIA 8.1.53 \cite{pythia} is used for W+jets cross section. The used values are listed in Tab. \ref{bxsec}. It should be noted that in any event analysis with fully hadronic final state, QCD multi-jet events may be an important background. As is seen in the next section, they can be suppressed by dedicated selection cuts which include the top tagging selection, top and W mass windows, the $b$-tagging, the jet transverse energy thresholds, etc. At the current analysis, $4\times 10^7$ QCD multi-jet events were analyzed and no event survived.
\begin{table}[h]
\centering
\begin{tabular}{|c|c|c|c|c|}
\hline
Process& $t\bar{t}$& Single top (s-channel) & Single top (t-channel) & W+jets \\
\hline
Cross section [pb] & 834 & 10.5 & 247 & 1.7$\times 10^{5}$\\
\hline
\end{tabular}
\caption{Background cross sections. \label{bxsec}}
\end{table}

\section{Event Generation and Analysis}
Signal events are generated using CompHEP in LHEF format (Les Houches Event File) \cite{lhef}. The output files are then passed to PYTHIA for further processing including final state showering, multiple interactions, decays, etc. Events are generated with charged Higgs mass hypotheses ranging from 400 GeV to 1 TeV with increments of 100 GeV. Background events are all generated using PYTHIA.\\
Since both signal and background processes are initiated from partons, parton distribution functions (PDF's) are used with a link between LHAPDF 5.8.6 \cite{lhapdf} and PYTHIA. In this analysis CTEQ 6.6 \cite{cteq} is used as the PDF set.

The jet reconstruction is performed using FASTJET 2.4.1 \cite{fastjet1,fastjet2} which is a multi-purpose jet reconstruction package.  A study of different jet reconstruction algoritms in the context of top tagging at LHC \cite{CA} has shown that the Cambridge-Aachen algorithm (CA) is better than $k_T$ \cite{kt} and anti-$k_T$ \cite{antikt} in terms of fake rate suppression and signal from background suppression. Therefore in this analysis, CA algorithm is used as the standard jet reconstruction algorithm.

The strategy of jet reconstruction is similar to what has been adopted in \cite{jss8} in the sense that a fat jet cone of $\Delta R = 1.5$ is used for top jet tagging, while for other jets outside the top decay cone, a standard jet cone with $\Delta R = 0.4$ is applied. Therefore the analysis uses a ``hybrid'' jet reconstruction and for each event, the jet reconstruction algorithm is executed twice: once for top tagging, and once for other jets independently.

The top tagging algorithm which is used in this analysis is HepTopTagger \cite{HEPTopTagger1} with a fat jet cone size of $\Delta  R=1.5$ and a jet $p_{T}$ threshold set to $p_T>150$ GeV. Signal events are in fully hadronic final state with a top jet (consisting of two light jets and a $b$-jet in the cone) and a $b$-jet outside the top decay cone. The HEPTopTagger procedure is described as follows.

The last step of C/A clustering for reconstruction of fat jets is reversed and the jet $j$ is decomposed into subjets $j_1$ and $j_2$ ($m_{j_1}$>$m_{j_2}$). If the subjets satisfy mass drop criterion $\min m_{j_i} < 0.8 \, m_j$, both subjets $j_1$ and $j_2$ are kept and decomposing continues for both. If the criterion fails, the subjet $j_2$ is discarded and decomposing continues only for the subjet $j_1$. Throughout this process, each subjet is decomposed if its mass is greater than 30 GeV and it has more than one constituent. Following this method, finally we reach a point, at which the decomposition stops completely and the procedure yields some subjets. In case of finding less than three subjets, the HEPTopTagger fails.

Having subjets in hand, a list of all their possible triplet combinations is created. The constituents of each triplet are now reclustered with $R_{filt}=min(0.3,\Delta R_{ij}/2)$, where $i$ and $j$ are the closest subjets. Reclustering yields a number of subjets, of which the five highest transverse momentum ones are kept as filtered subjets, and all others are discarded. This filtering is performed to clean jets from contamination due to the underlying events and pile-up. The filtered mass is defined to be the invariant mass of the five kept filtered subjets. Following this procedure, a set of filtered subjets is obtained for each triplet combination. The set of filtered subjets for which the filtered mass is closest to the top quark mass is kept and all other sets are discarded. The constituents of the kept set of filtered subjets are now again reclustered by exclusive C/A algorithm, which causes the jet to have exactly three subjets $j_1$, $j_2$ and $j_3$ ($p_{{\mathrm{T},1}}>p_{{\mathrm{T},2}}>p_{{\mathrm{T},3}}$).

At this point, a mass selection is performed on the final three subjets. The invariant mass $m_{123}$ of the three subjets must be in the top mass window, and the invariant masses $m_{12}$, $m_{23}$ and $m_{13}$ associated with various pairs of the subjets, must satisfy at least one of the following three conditions:
\begin{alignat}{5}
&0.2 <\arctan \frac{m_{13}}{m_{12}} < 1.3
\qquad \text{and} \quad
R_{\min}< \frac{m_{23}}{m_{123}} < R_{\max}
\notag \\
&R_{\min}^2 \left(1+\left(\frac{m_{13}}{m_{12}}\right)^2 \right)
< 1-\left(\frac{m_{23}}{m_{123}} \right)^2
< R_{\max}^2 \left(1+\left(\frac{m_{13}}{m_{12}}\right)^2 \right)
\quad \text{and} \quad
\frac{m_{23}}{m_{123}} > R_\text{soft}
\notag \\
&R_{\min}^2\left(1+\left(\frac{m_{12}}{m_{13}}\right)^2 \right)
< 1-\left(\frac{m_{23}}{m_{123}} \right)^2
< R_{\max}^2\left(1+\left(\frac{m_{12}}{m_{13}}\right)^2 \right)
\quad \text{and} \quad
\frac{m_{23}}{m_{123}}> R_\text{soft}
\label{eq:heptop}
\end{alignat}
Where the dimensionless mass windows $R_{\min}=85\% \times m_W/m_t$ and
$R_{\max}=115\% \times m_W/m_t$, and the soft cutoff $R_\text{soft} =
0.35$ removes QCD events which are not identified as soft radiation by the C/A algorithm. The passed subjets are idetified as top quark subjets if they satisfy one another condition. The overall transverse momentum of the three subjets must be greater than 150 GeV, since the decay jets of the top quark are assumed to be merged in one jet with radius $R=1.5$. Having passed this final check, the top quark identification is completed and the subjets are tagged as top quark subjets.

\begin{figure}
 \centering \includegraphics[width=0.7\textwidth]{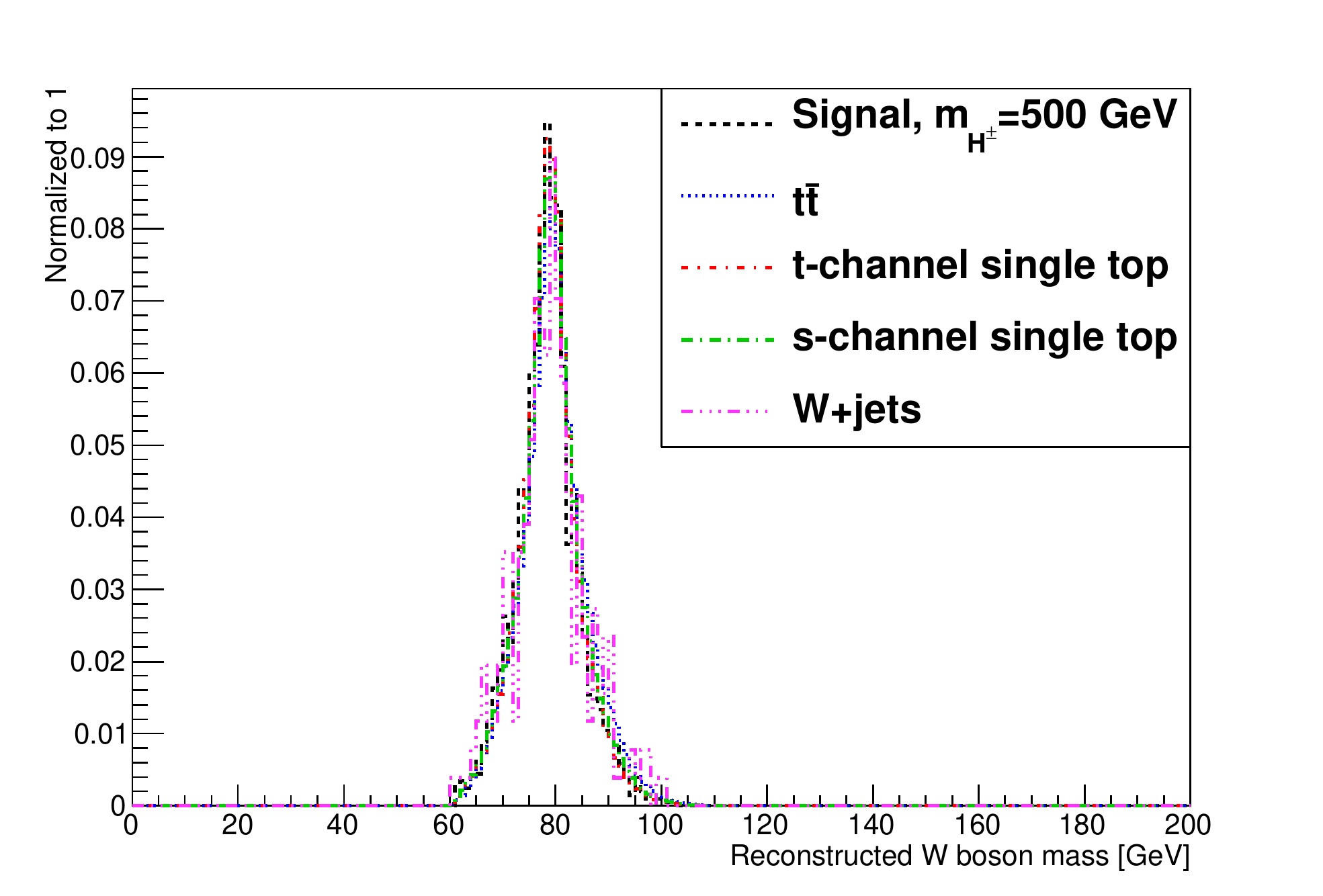}
 \caption{The reconstructed W boson mass obtained from the top tagging algorithm. \label{wmass}}
 \end{figure}

\begin{figure}
 \centering \includegraphics[width=0.7\textwidth]{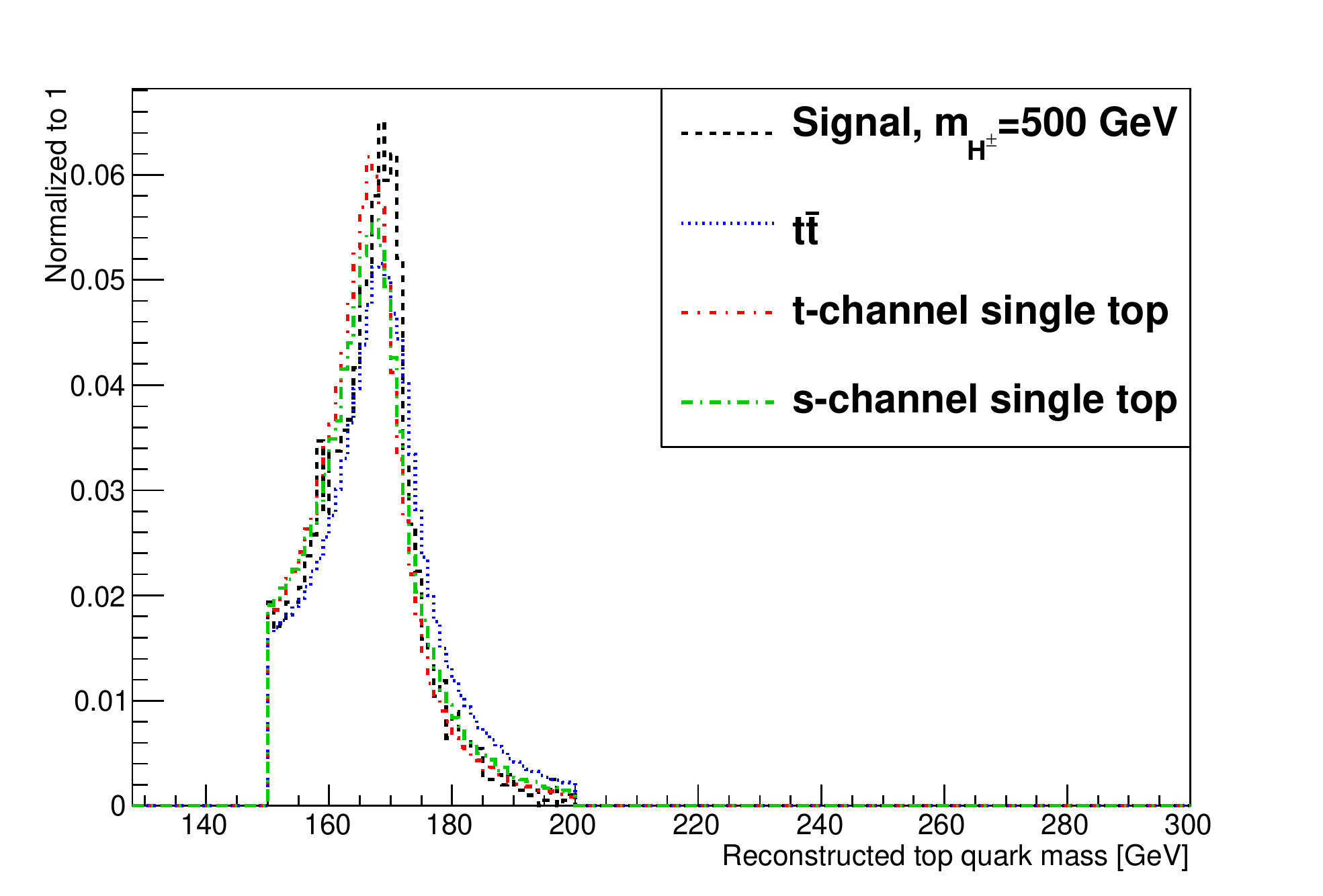}
 \caption{The reconstructed top quark mass obtained from the top tagging algorithm. \label{topmass}}
 \end{figure}

The final result of the top tagging algorithm is accepted if all three jets satisfy the top mass requirement as in Eq. \ref{masswindow}. Figures \ref{wmass} and \ref{topmass} show the distribution of the reconstructed W boson and top quarks in signal and background events.
\be
150\textnormal{~GeV}<m_{j1j2b}<190\textnormal{~GeV}
\label{masswindow}
\ee

In the next step a search for standard jets is performed and number of jets outside the top decay cone is counted. The kinematic threshold for jets is as in Eq. \ref{jetet}.
\be
\textnormal{jet~}E_{T} > 150 \textnormal{~GeV},~~|\eta|<5
\label{jetet}
\ee

\begin{figure}
 \centering \includegraphics[width=0.7\textwidth]{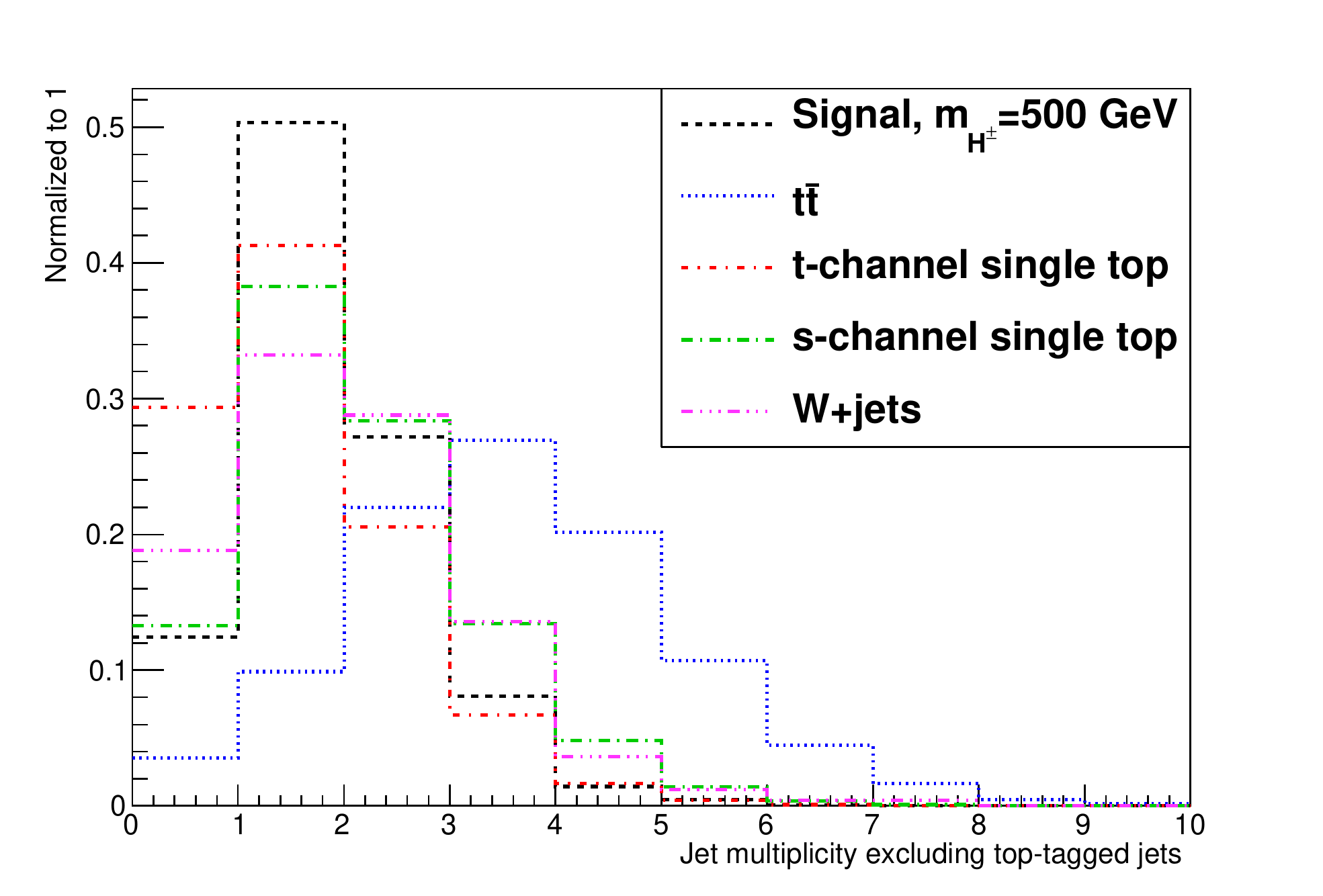}
 \caption{Jet multiplicity excluding jets from the top quark decay. \label{jetmul}}
 \end{figure}

\begin{figure}
 \centering \includegraphics[width=0.7\textwidth]{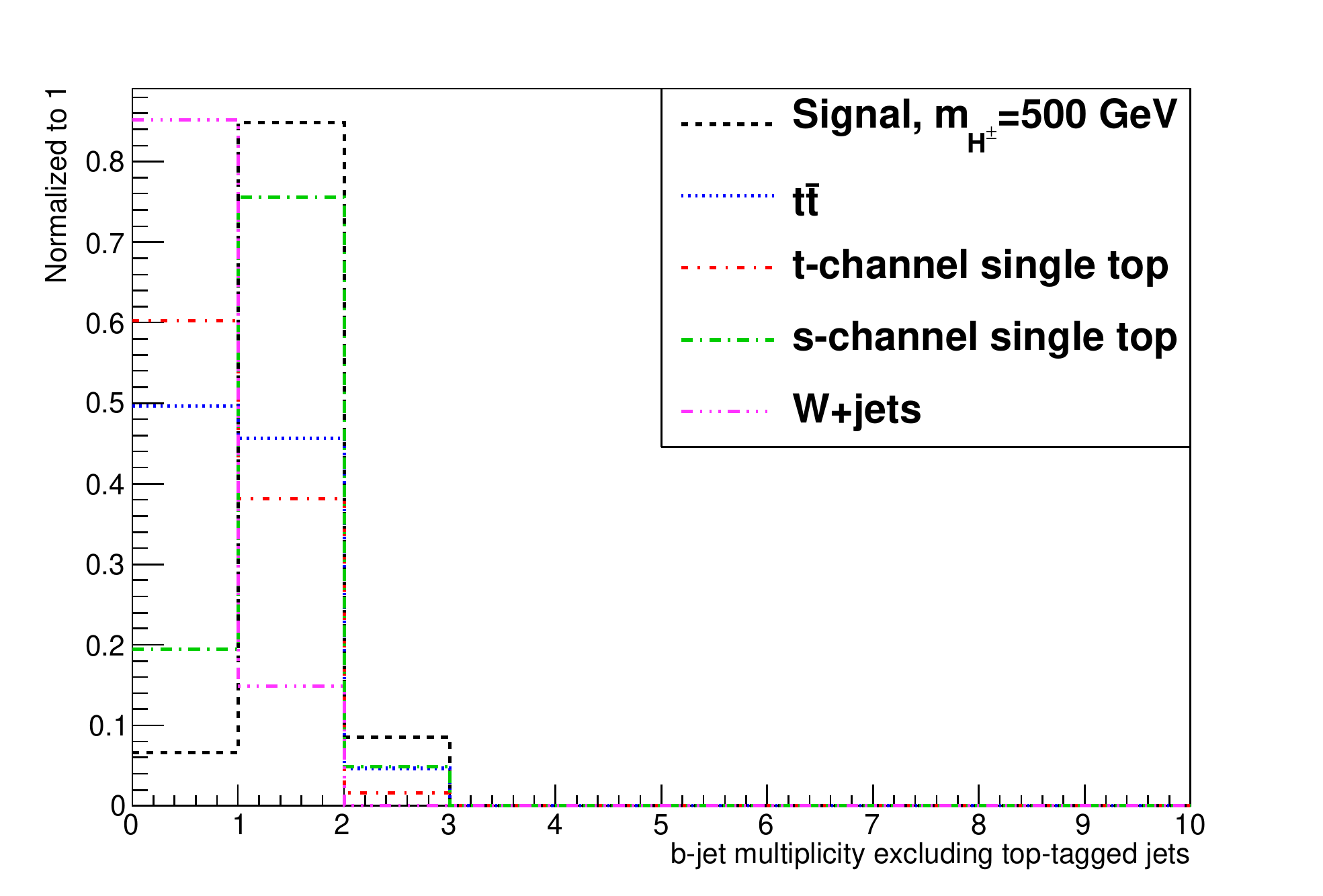}
 \caption{B-jet multiplicity excluding jets from the top quark decay. \label{bjetmul}}
 \end{figure}

Figure \ref{jetmul} shows the total number of jets in signal and background events excluding those which satisfy the top tagging requirements and fall inside the top decay cone. The $b$-tagging is then applied on selected jets in the previous step. For each jet, a search is performed among adjacent $b$ or $c$ quarks using generator level information. A jet is selected as a $b$-jet with 60$\%$(10$\%$) probability if it is near a $b$($c$) quark. The algorithm thus uses the typical $b$-tagging efficiency of LHC experiments \cite{btag}. Figure \ref{bjetmul} shows number of $b$-jets outside the top decay cone. An event is selected if there is only one $b$-jet outside the top tagging cone.

\begin{figure}
 \centering \includegraphics[width=0.7\textwidth]{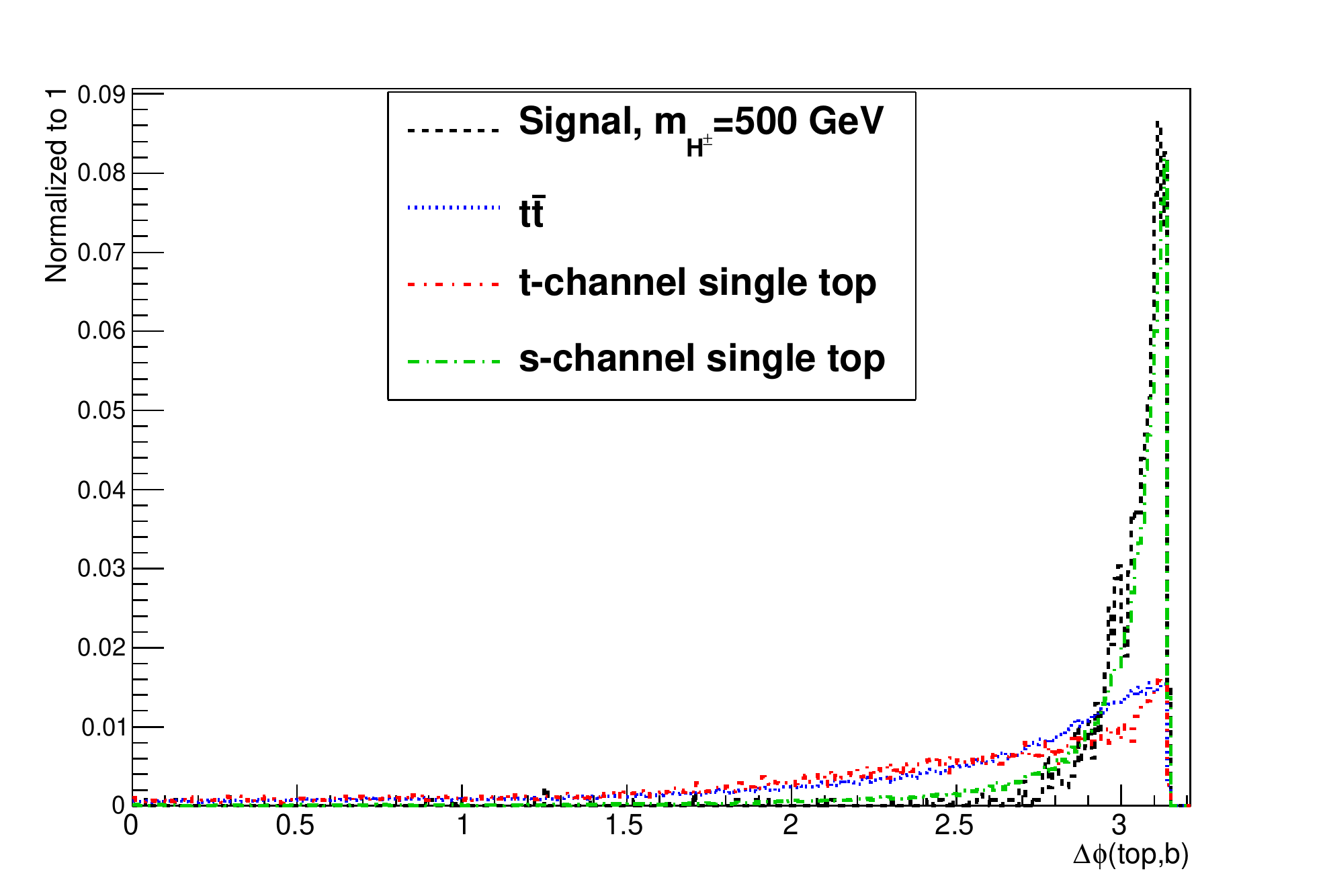}
 \caption{The azimuthal angle difference between the top and bottom quarks. \label{dphi} }
 \end{figure}

Another aspect of signal events is that they tend to produce the top and bottom quark pair in opposite directions due to the nature of $s$-channel processes. This feature should appear in azimuthal plane as well. Therefore the angle between the top quark and bottom quark (outside the top decay cone) is calculated and the result is plotted as a distribution of $\Delta\phi$ for both signal and background events for comparison as seen in Fig. \ref{dphi}. According to Fig. \ref{dphi}, a selection cut is applied as in Eq. \ref{dphieq}.
\be
\Delta\phi_{\textnormal{(top quark, bottom quark)}} > 2.8
\label{dphieq}
\ee
When all selection cuts are applied, a chain of selection efficiencies is obtained as in Tab. \ref{sseleff} and Tab. \ref{bseleff}.
\begin{table}[h]
\centering
\begin{tabular}{|c|c|c|c|c|c|c|c|}
\hline
$m_{H^{\pm}}$ [GeV]& 400 & 500 & 600 & 700 & 800 & 900 & 1000\\
\hline
One tagged top & 0.015 & 0.097 & 0.19 & 0.25 & 0.29 & 0.32 & 0.34\\
\hline
One jet & 0.37 & 0.85 & 0.92 & 0.94 & 0.95 & 0.95 & 0.95\\
\hline
$b$-tagging & 0.9 & 0.97 & 0.98 & 0.97 & 0.97 & 0.98 & 0.98\\
\hline
$\Delta \phi(top,bottom)$ & 0.92 & 0.97 & 0.97 & 0.98 & 0.99 & 0.99 & 0.99\\
\hline
Total eff. & 0.0047 & 0.077 & 0.16 & 0.22 & 0.27 & 0.3 & 0.31\\
\hline
\end{tabular}
\caption{Signal selection efficiencies assuming different charged Higgs masses. \label{sseleff}}
\end{table}

\begin{figure}
 \centering \includegraphics[width=0.7\textwidth]{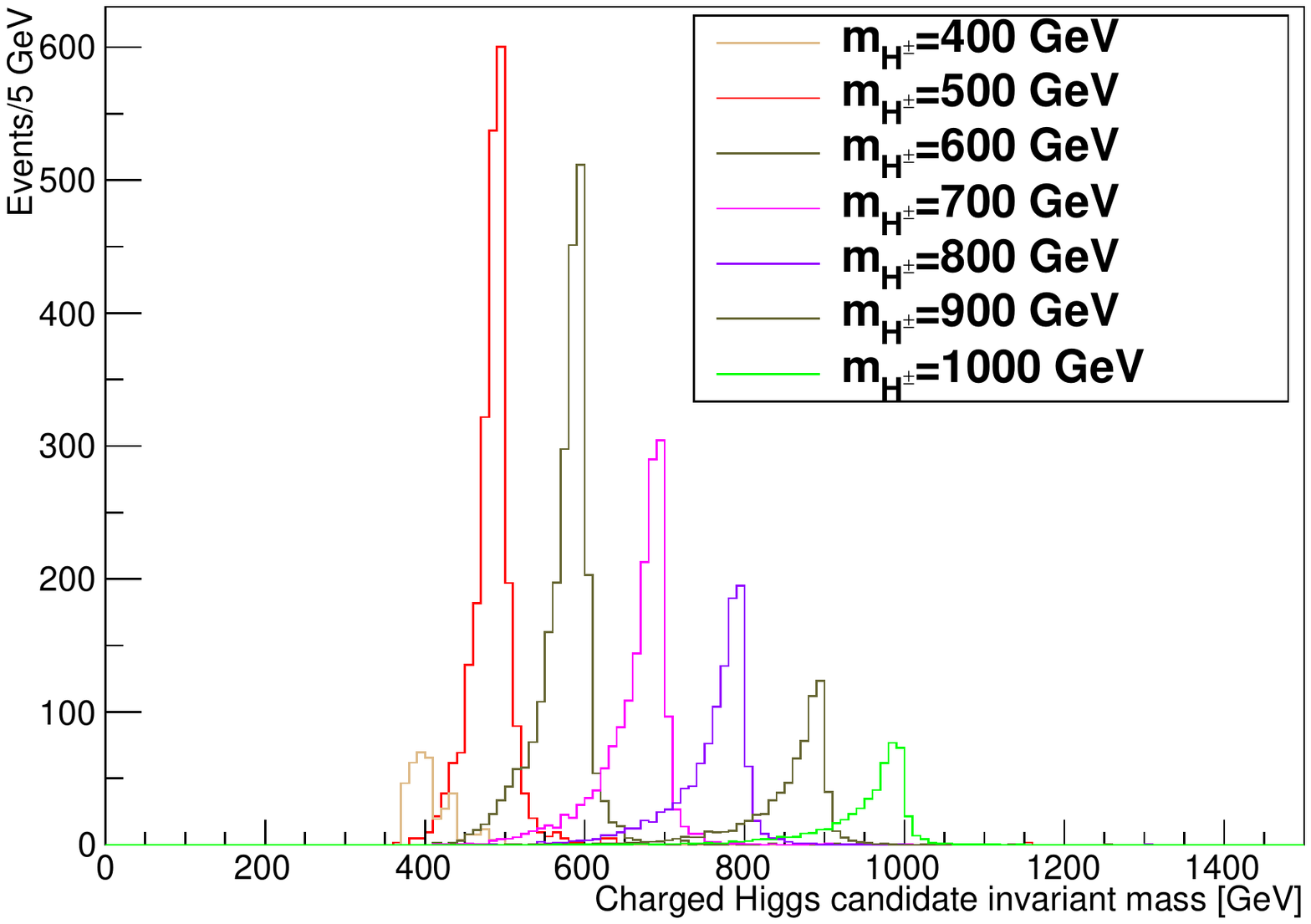}
 \caption{The reconstructed charged Higgs mass in signal events for different charged Higgs mass hypotheses. \label{allsignal}}
 \end{figure}

\begin{table}[h]
\centering
\begin{tabular}{|c|c|c|c|c|c|}
\hline
Process& $t\bar{t}$& Single top (s-channel) & Single top (t-channel) & W+jets & QCD \\
\hline
One tagged top & 0.13 & 0.023 & 0.017 & 4.9e-05 & 5.8e-06\\
\hline
One jet & 0.47 & 0.73 & 0.62 & 0.62 & 0.61\\
\hline
$b$-tagging  & 0.21 & 0.86 & 0.055 & 0.0099 & 0.007\\
\hline
$\Delta \phi(top,bottom)$ & 0.73 & 0.91 & 0.82 & 1 & 0\\
\hline
Total eff. & 0.0099 & 0.013 & 0.00047 & 3e-07 & 0\\
\hline
\end{tabular}
\caption{Background selection efficiencies. \label{bseleff}}
\end{table}
In signal events the top and bottom quark come from a charged Higgs boson and their invariant mass should be in principle make the charged Higgs boson mass. However due to jet energy resolution, mis-identification of jets and errors in their energy and flight directions, and false jet combination, a distribution of invariant mass with a peak at (almost) the nominal charged Higgs mass is obtained. This distribution is seen in Fig. \ref{allsignal} where different charged Higgs mass hypotheses are tested in the simulation. The normalization is based on total selection efficiencies (Tab. \ref{sseleff}) and cross sections (Fig. \ref{Xsecs}) and the integrated luminosity according to $\sigma \times \epsilon \times L$ where $\sigma$ is the signal cross section, $\epsilon$ is the total efficiency and $L$ is the integrated luminosity.

\begin{figure}
  \centering \includegraphics[width=0.7\textwidth]{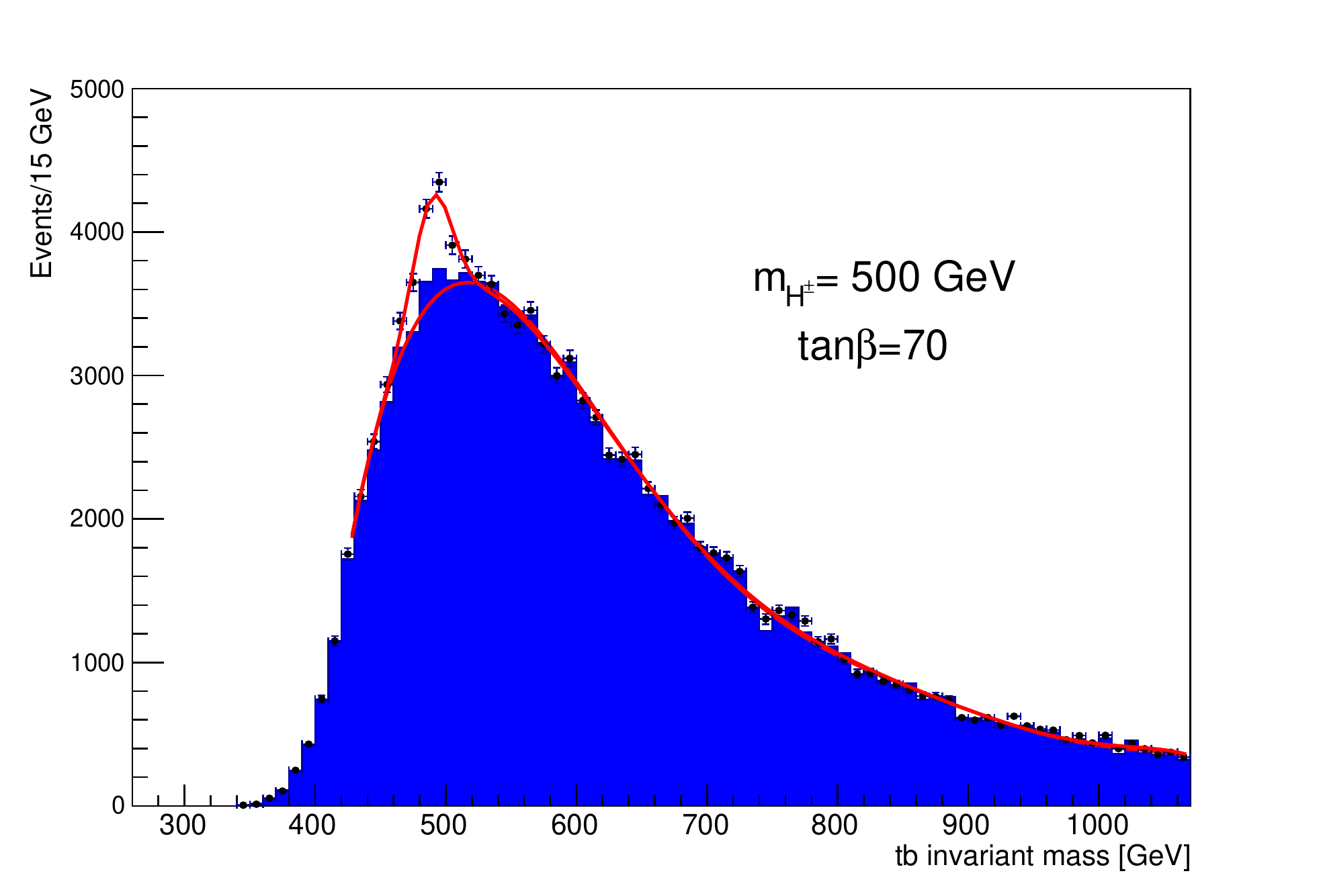}
 \caption{Charged Higgs signal on top of the total background assuming $m_{H^{\pm}}=500$ GeV. The pseudo-data are shown with statistical error bars. The background fit and the fit to signal plus background are also shown as solid lines. \label{500}}
 \end{figure}

\begin{figure}
  \centering \includegraphics[width=0.7\textwidth]{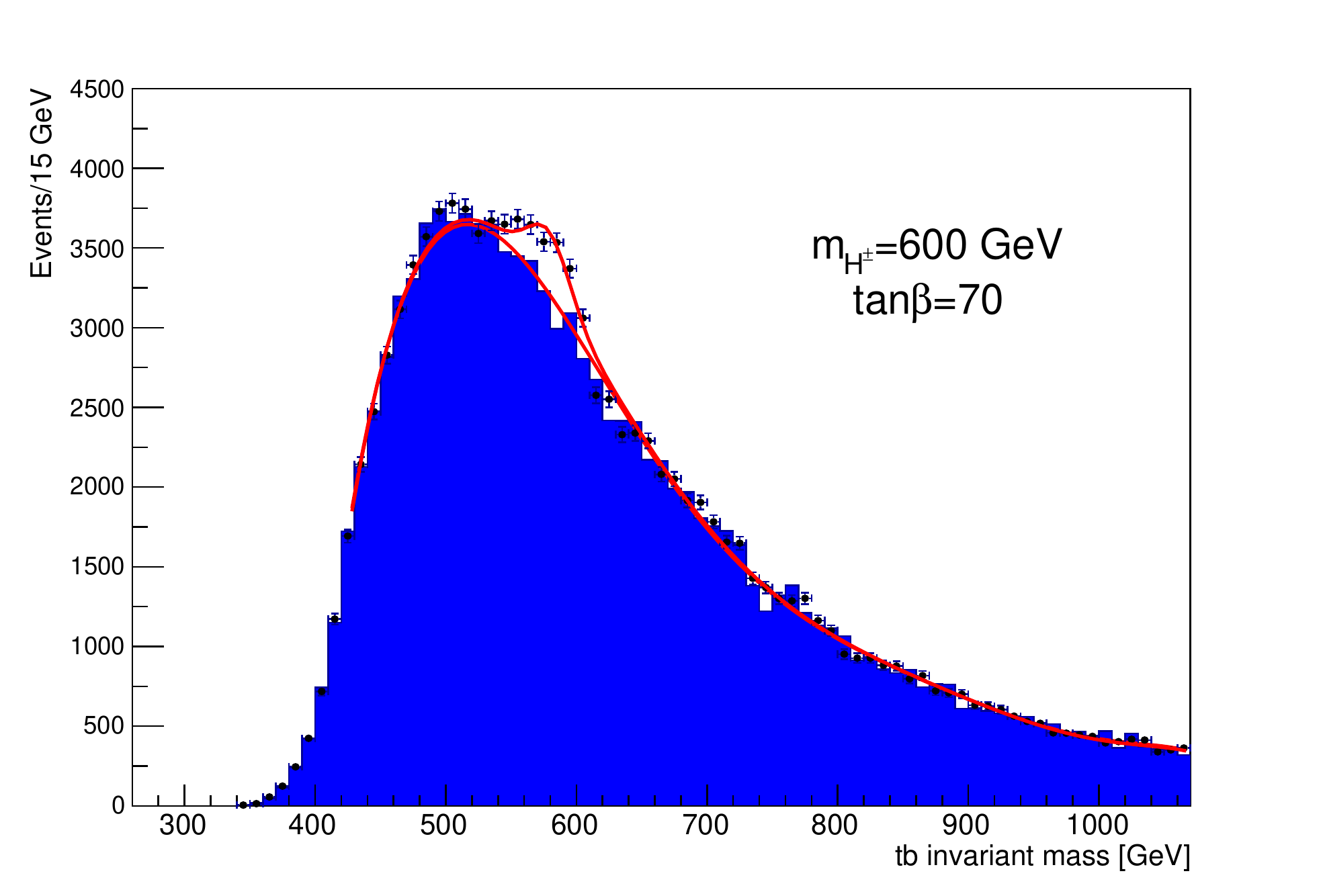}
 \caption{Charged Higgs signal on top of the total background assuming $m_{H^{\pm}}=600$ GeV. The pseudo-data are shown with statistical error bars. The background fit and the fit to signal plus background are also shown as solid lines. \label{600}}
 \end{figure}

\begin{figure}
  \centering \includegraphics[width=0.7\textwidth]{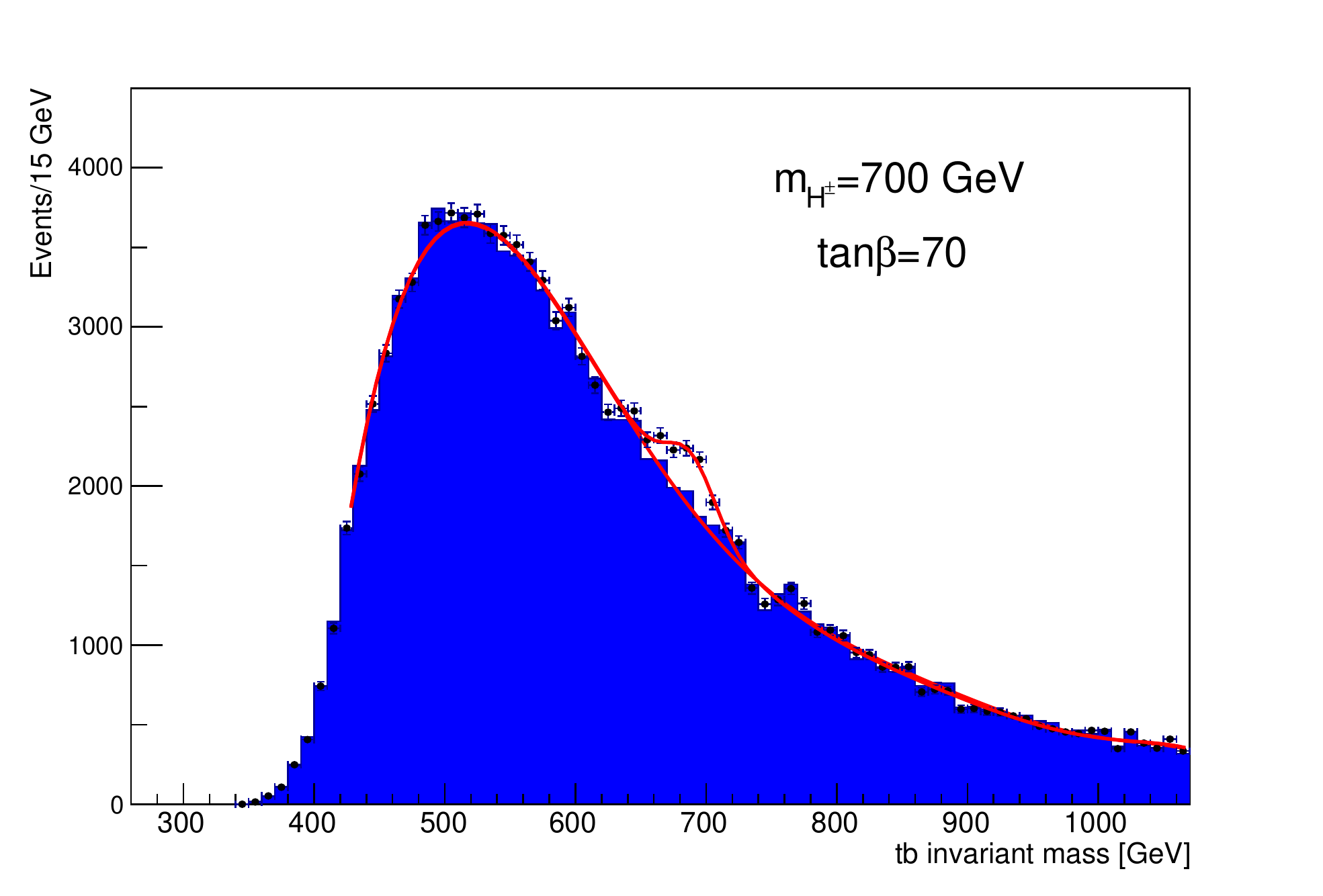}
 \caption{Charged Higgs signal on top of the total background assuming $m_{H^{\pm}}=700$ GeV. The pseudo-data are shown with statistical error bars. The background fit and the fit to signal plus background are also shown as solid lines. \label{700}}
 \end{figure}

\begin{figure}
  \centering \includegraphics[width=0.7\textwidth]{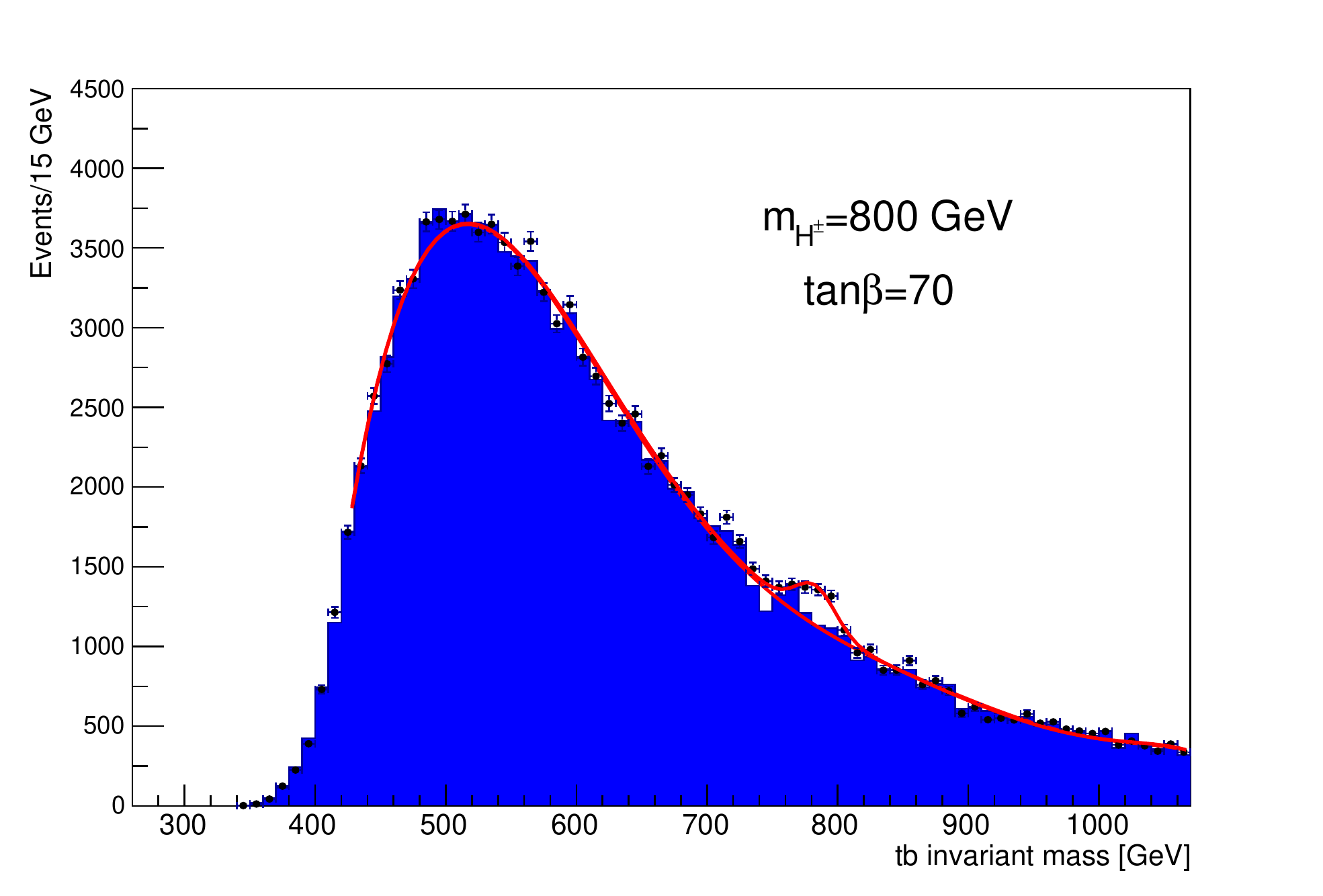}
 \caption{Charged Higgs signal on top of the total background assuming $m_{H^{\pm}}=800$ GeV. The pseudo-data are shown with statistical error bars. The background fit and the fit to signal plus background are also shown as solid lines. \label{800}}
 \end{figure}

\begin{figure}
  \centering \includegraphics[width=0.7\textwidth]{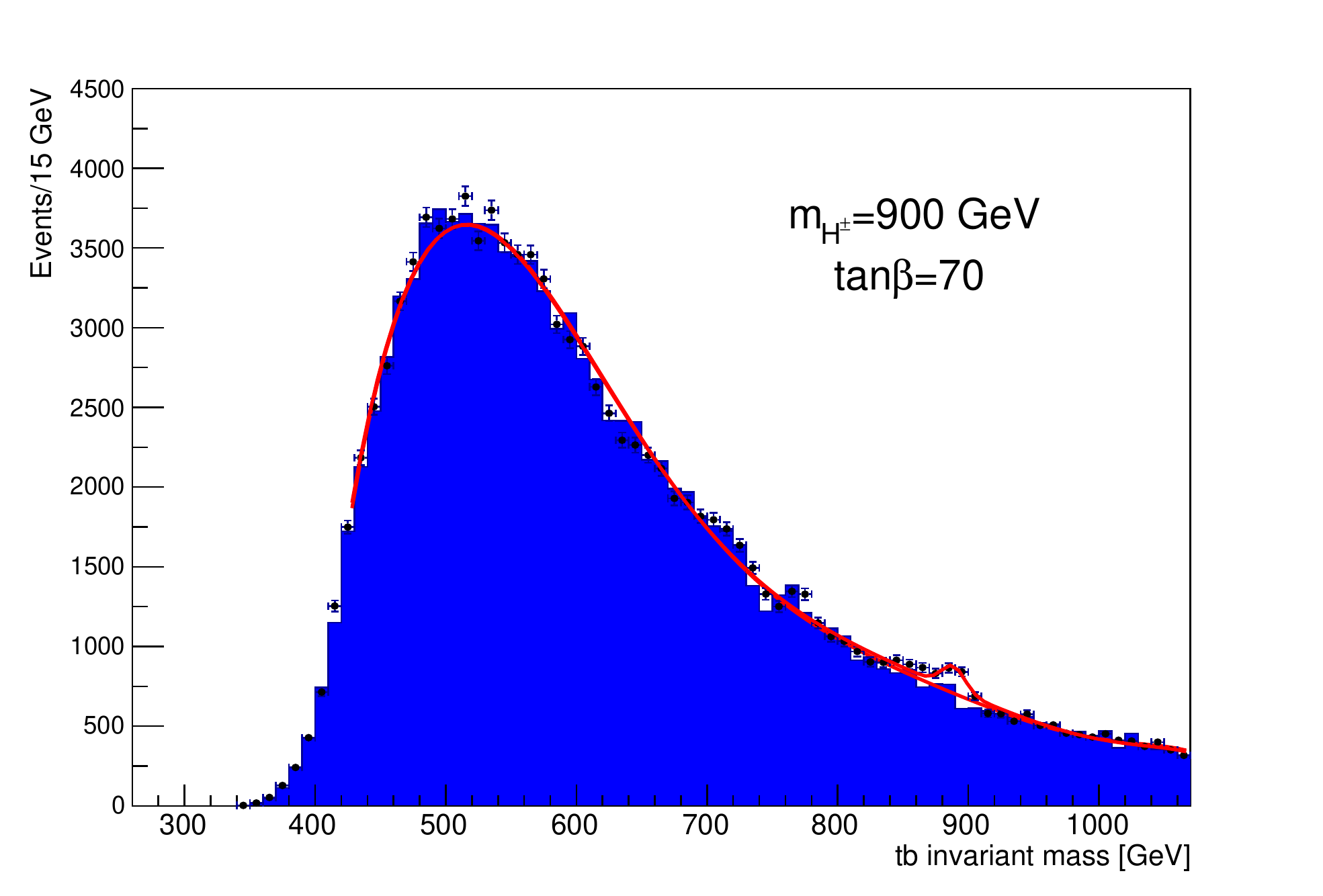}
 \caption{Charged Higgs signal on top of the total background assuming $m_{H^{\pm}}=900$ GeV. The pseudo-data are shown with statistical error bars. The background fit and the fit to signal plus background are also shown as solid lines. \label{900}}
 \end{figure}

\begin{figure}
  \centering \includegraphics[width=0.7\textwidth]{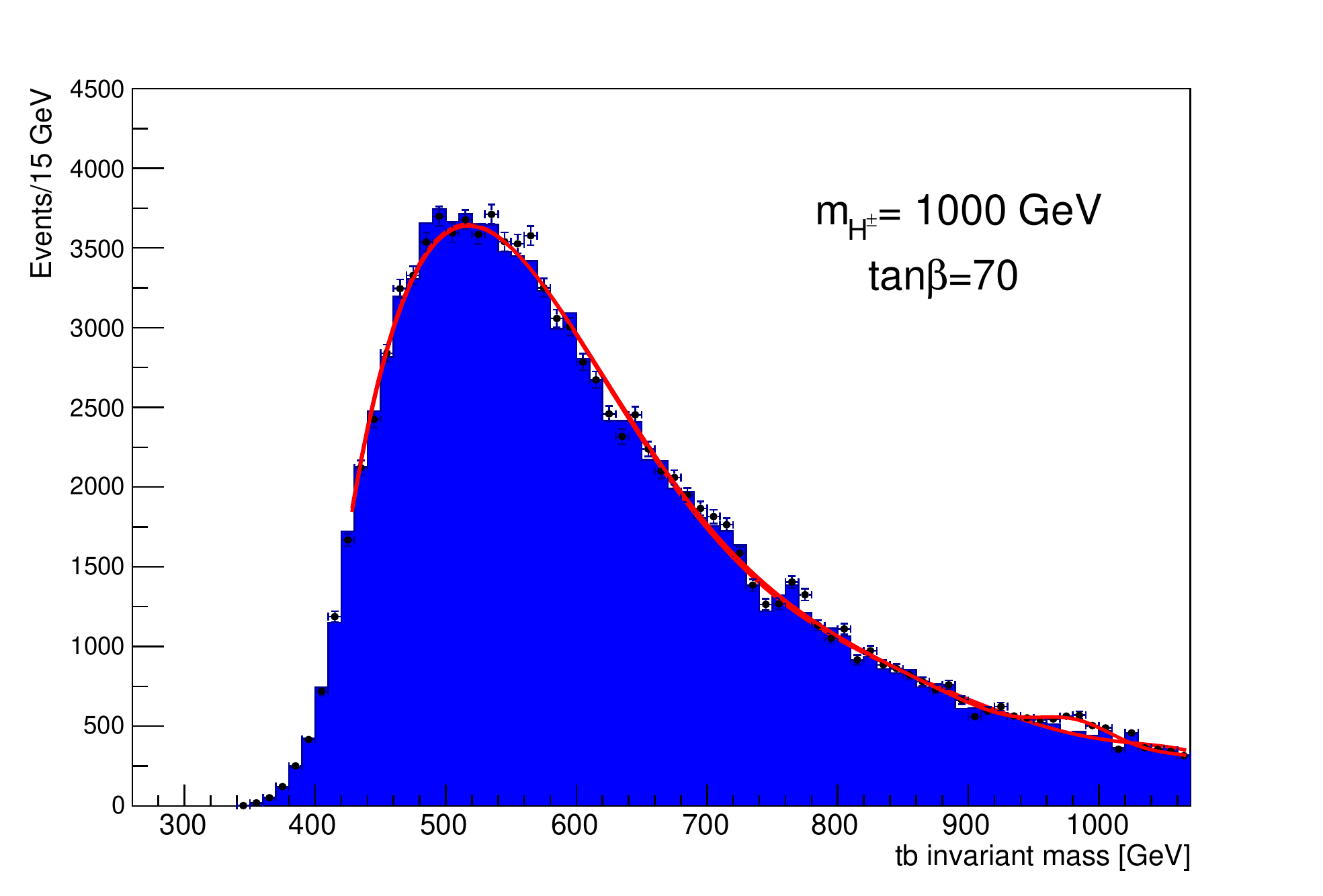}
 \caption{Charged Higgs signal on top of the total background assuming $m_{H^{\pm}}=1000$ GeV. The pseudo-data are shown with statistical error bars. The background fit and the fit to signal plus background are also shown as solid lines. \label{1000}}
 \end{figure}

Figure \ref{sb} shows signal distributions on top of the total SM background. Due to the large background cross section, signal events appear as small excess of events on top of the background centered at the assumed charged Higgs mass. The natural question is now whether the charged Higgs mass can be extracted from a fit to signal plus background distribution. In order to obtain the answer, RooFit package inside ROOT 5.34 \cite{root} is used to simulate toy experiment using probability density distributions taken from signal and background shapes of the charged Higgs candidate invariant mass. In this simulation, for each charged Higgs mass assumption, several toy experiments (LHC's) were performed to check the possibility of the fit to signal and background distributions. Figures \ref{500}-\ref{1000} show a typical (selected) experiment demonstrating total background, simulated data with statistical error bars using the PDF's of Figs. \ref{allsignal} and \ref{sb} and the fit to the total distribution. The fit function uses a polinomial and a gaussian function. As seen the signal is well distinguished from the background for charged Higgs masses ranging from 500 to 1000 GeV. The 400 GeV charged Higgs could not be well distinguished as it lies on a large and increasing background.

\begin{figure}
  \centering \includegraphics[width=0.7\textwidth]{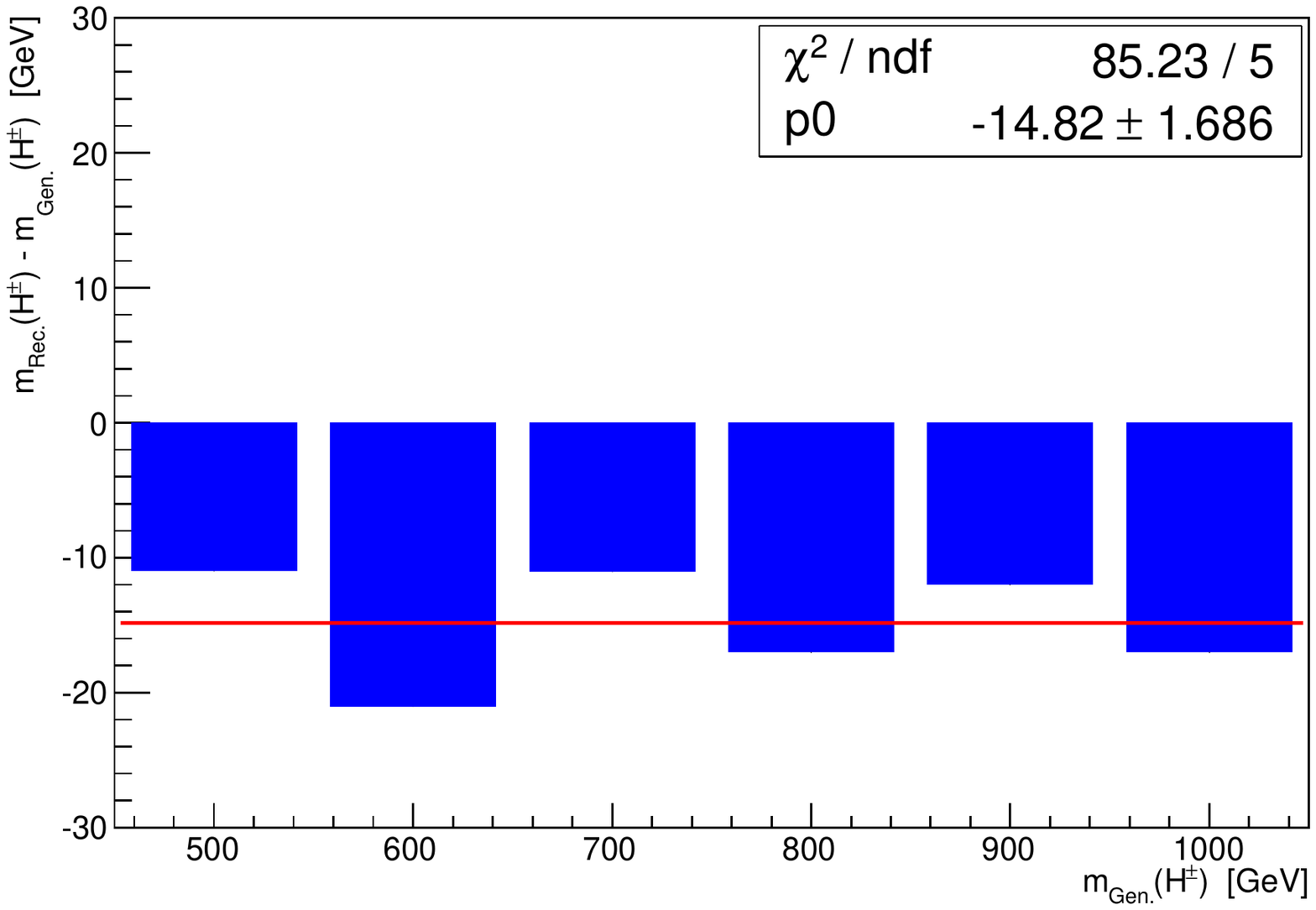}
 \caption{Reconstructed charged Higgs mass minus generated value. \label{deltam}}
 \end{figure}

The parameters of the Gaussian part of the fit function can be used to determine the charged Higgs mass. The difference between the reconstructed and generated charged Higgs masses is plotted in Fig. \ref{deltam}.

As seen in Fig. \ref{deltam}, there is an overall negative shift in the mean value of the Gaussian fit with respect to the generated charged Higgs mass. The reason for this should be due to the jet reconstruction algorithm: its parameters (cone size, seed threshold, ...) and the uncertainties arised from them. Since the charged Higgs invariant mass is reconstructed in a fully hadronic environment, the only source of uncertainty should be related to jets. A detailed study of jet four momentum and matching it with the MC truth (i.e. the jet original particle four momentum) could reveal the actual reason for the observed offset. One may also tune the jet algorithm parameters such as the cone size to contain all jet activity inside.

\begin{figure}
 \centering \includegraphics[width=0.7\textwidth]{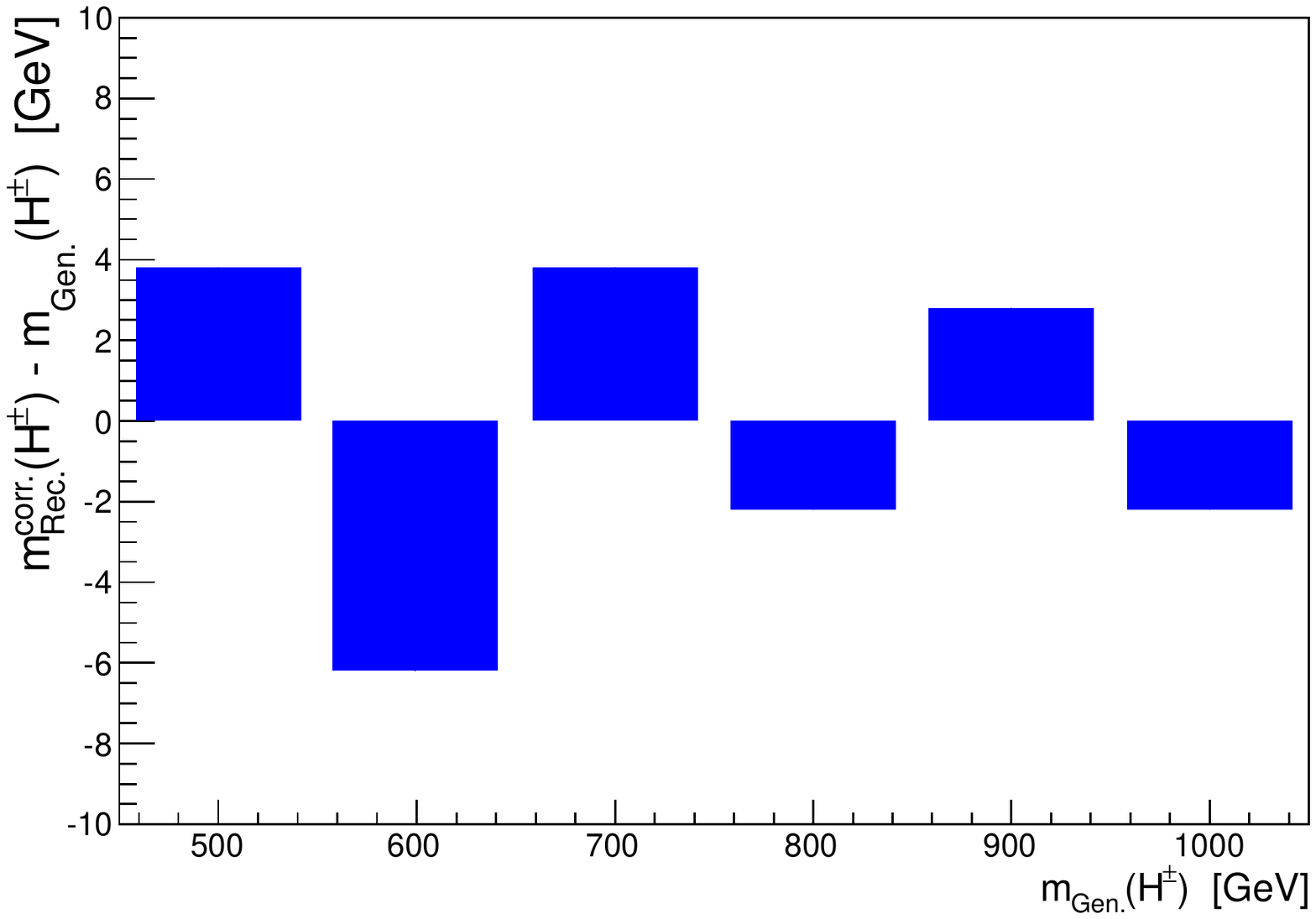}
 \caption{Reconstructed charged Higgs mass minus generated value after correction. \label{deltam_corr}}
 \end{figure}

In a real experiment, all sources of uncertainties are taken into account including electronic noise, pile-up, etc. A final correction to the jet four momentum may include several multiplicative factors for off-set effects, Data/MC calibration and jet energy scale uncertainties \cite{jes}. Since a study of this type is beyond the scope of this analysis, a simple off-set correction is applied to match reconstructed charged Higgs invariant mass with its true value. In order to do this, a flat function is used to fit to the plot of Fig. \ref{deltam} and obtain the average distance of the reconstructed mass from the generated mass. The average value is -14.8 GeV. Therefore all reconstructed masses are increased by 14.8 GeV. This correction leads to Fig. \ref{deltam_corr} where the difference between the corrected reconstructed masses and generated masses is plotted.

The corrected charged Higgs masses are listed in Tab. \ref{masses} including fit uncertainties. The charged Higgs mass can thus be measured with few GeV uncertainty. This is however a statistical error. In a real experiment, there are systematic uncertainties which should be taken into account. Since this is a fully hadronic final state analysis, the dominant uncertainty should come from jet energy uncertainty which is expected to be less than $1\%$ in the central region of detector for jet transverse energies in the range 55 to 500 GeV at the current LHC stage \cite{jes}. Other sources of uncertainties include the $b$-tagging uncertainty, the uncertainty from the fit function used in the analysis and the background modeling in making the total background probability density function ($pdf$). The latter is an important part of the analysis which relies on a correct understanding of the background distributions. This is well achievable in a real data analysis where distributions of different background samples taken from real data and MC are compared to obtain a reasonable $pdf$ of the total background.
\begin{table}
\centering
\begin{tabular}{|c|c|c|c|c|c|c|}
\hline
Generated mass [GeV] & 500 & 600 & 700 & 800 & 900 & 1000\\
\hline
Reconstructed mass [GeV] & 504 $\pm$1 & 594$\pm$3 & 704$\pm$2 & 798$\pm$2 & 903$\pm$3 & 998$\pm$3\\
\hline
\end{tabular}
\caption{Generated and reconstructed charged Higgs masses. The errors are statistical. \label{masses}}
\end{table}

\begin{table}
\centering
\begin{tabular}{|c|c|c|c|c|c|c|c|}
\hline
$m_{H^{\pm}}$ [GeV]& 400 & 500 & 600 & 700 & 800 & 900 & 1000\\
\hline
Mass window [GeV] & 380-400 & 480-500 & 570-610 & 670-700 & 770-800 & 870-910 & 960-1000\\
\hline
Total eff. & 0.0023 & 0.047 & 0.11 & 0.12 & 0.14 & 0.17 & 0.18 \\
\hline
S & 177 & 1459 & 1660 & 950 & 618 & 440 & 287 \\
\hline
B & 780 & 10702 & 15536 & 7927 & 4839 & 3491 & 2402 \\
\hline
$S/B$ &  0.23 & 0.14 & 0.11 & 0.12 & 0.13 &  0.13 & 0.12 \\
\hline
$S/\sqrt{B}$ &  6.3 & 14 & 13 & 11 & 8.9 & 7.4 & 5.8 \\
\hline
\end{tabular}
\caption{Charged Higgs mass window cuts, nutmber of signal and background events after all selection cuts and mass window cut and the final signal to background ratio and signal significance. \label{signif_table}}
\end{table}
\section{Signal Significance and Phase Space Contours}
\begin{figure}
 \centering \includegraphics[width=0.7\textwidth]{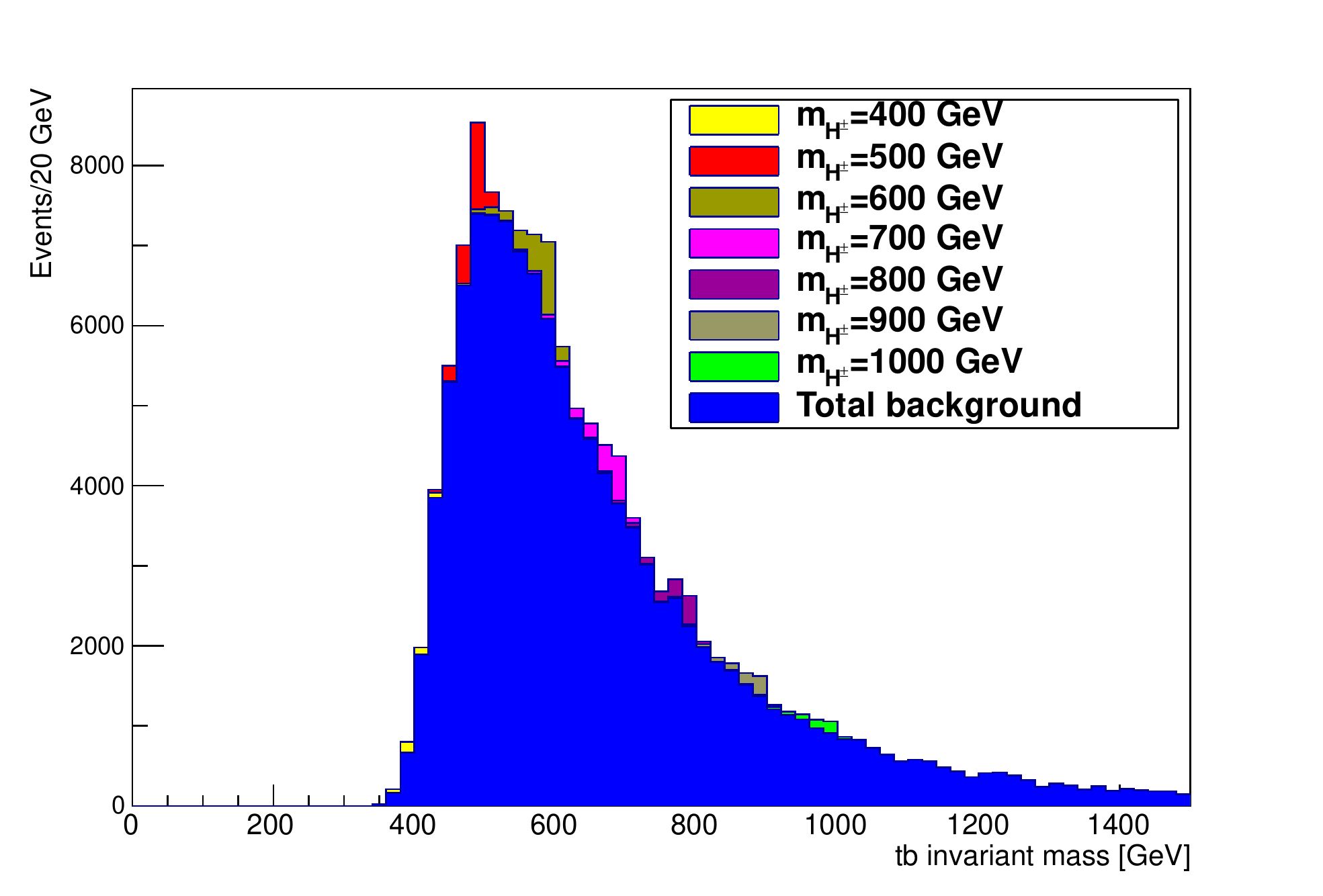}
 \caption{Distribution of signal plus background including different charged Higgs mass peaks independently. \label{sb}}
 \end{figure}

\begin{figure}
\centering \includegraphics[width=0.7\textwidth]{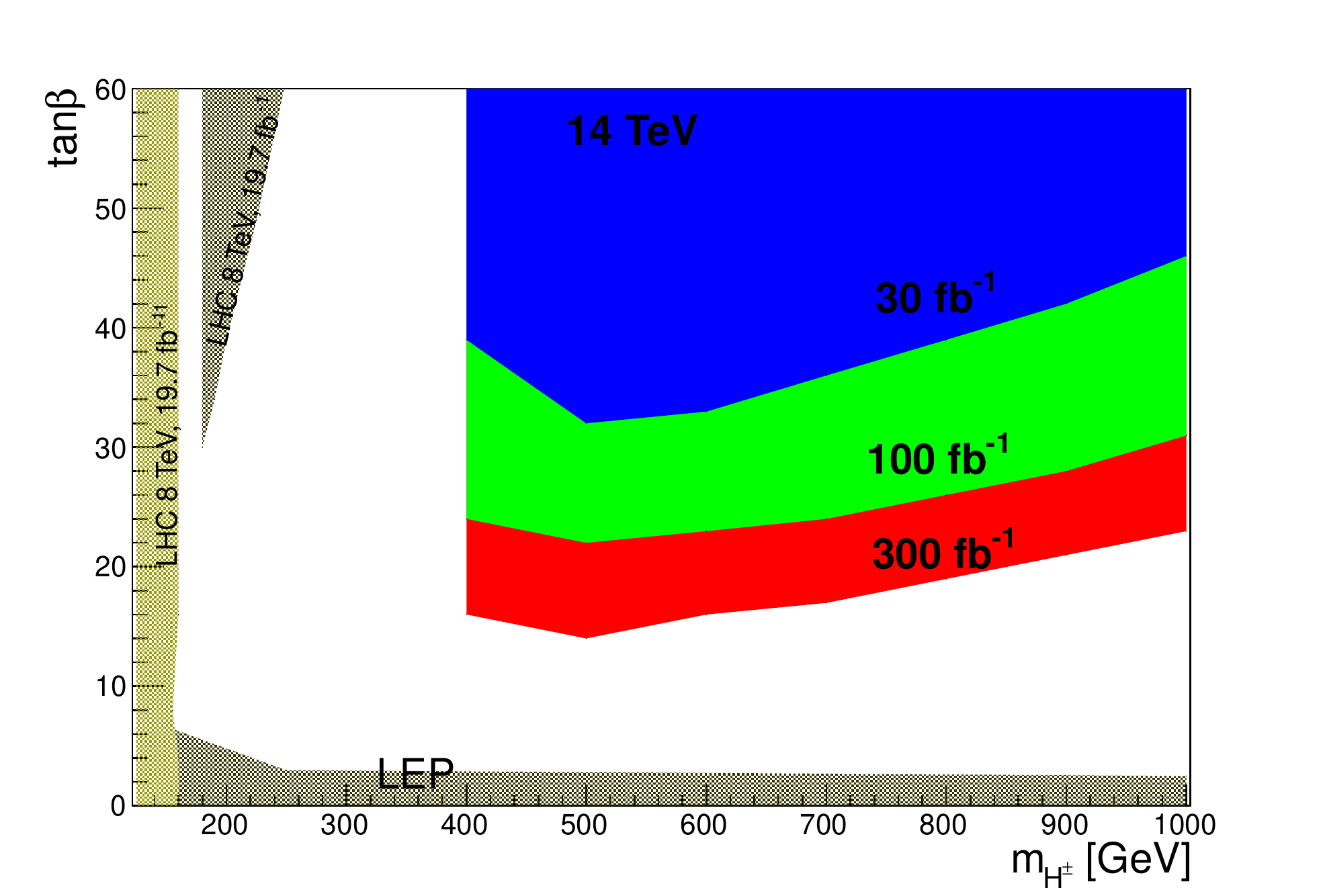}
 \caption{The 95$\%$ C.L. contour as a function of the charged Higgs mass and \tanb at different integrated luminosities. \label{95cl}}
\end{figure}

\begin{figure}
\centering \includegraphics[width=0.7\textwidth]{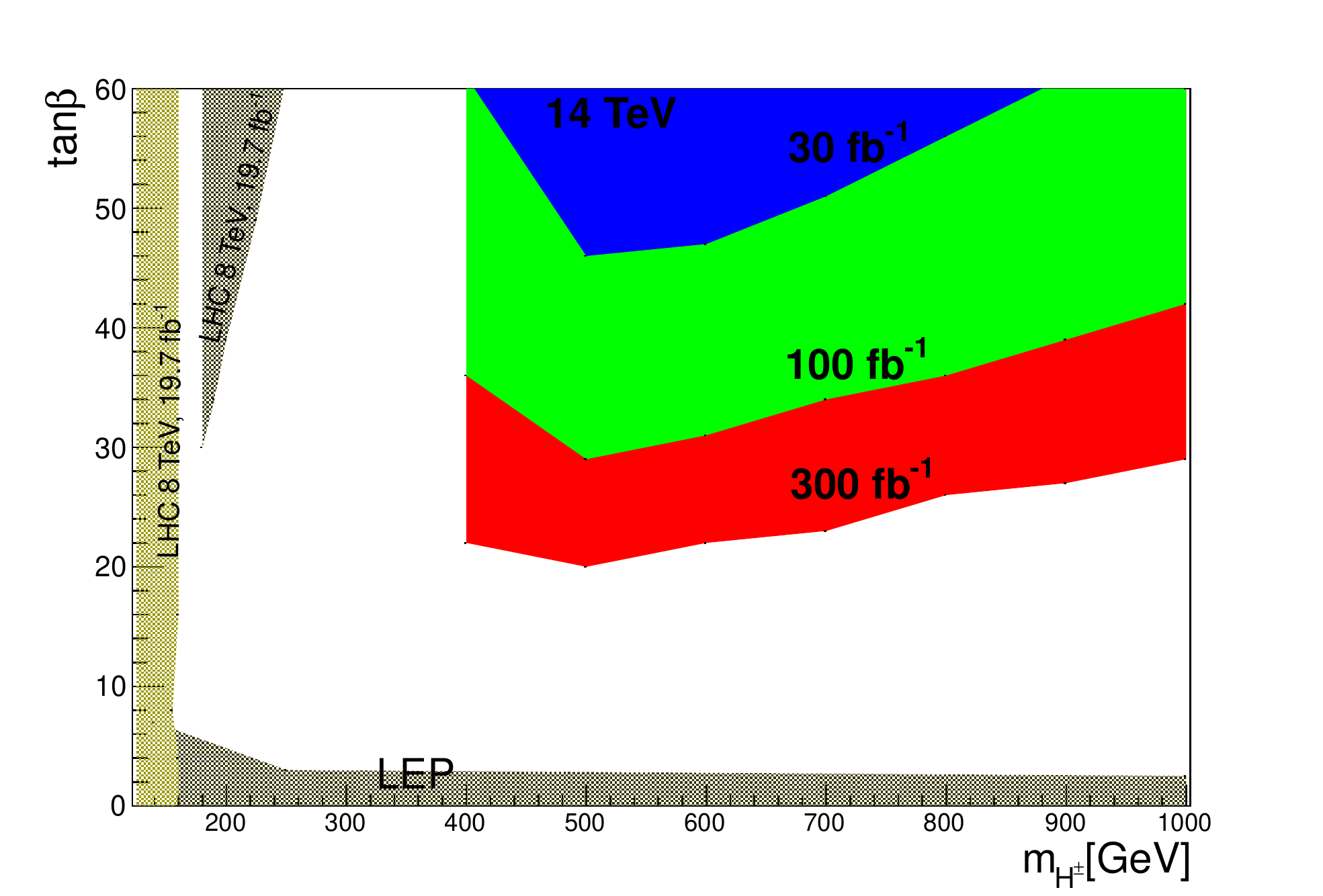}
\caption{The 5 $\sigma$ discovery contour as a function of the charged Higgs mass and \tanb at different integrated luminosities. \label{5sigma}}
\end{figure}

Using the signal plus background distributions of Fig. \ref{sb}, a mass window cut is applied to increase the signal significance. The cut window is optimized so as to give the highest significance. Table \ref{signif_table} shows mass window cuts, the total selection efficiency, signal to background ratio and signal significance at $L$=30 \invfb
and \tanb = 70. In order to obtain the signal significance at other points of $m_{H^{\pm}},\tan\beta$ phase space, it is assumed that variation of \tanb only changes the cross section due to the change in vertex couplings and does not change the kinematics of events or selection efficiencies. Therefore the signal cross section is calculated at each point of parameter space using CompHep, then TLimit code, which is implemented in ROOT, is used to obtain the signal significance. Results are finally quoted in terms of contours which show the minimum \tanb required to exclude a chaged Higgs with a given mass at 95$\%$ confidence level or discover at 5$\sigma$. These contours are shown in Fig. \ref{95cl} and \ref{5sigma} where different luminosities have been assumed. The current exclusion contours from LEP and LHC experiments are also included. As seen, LHC exclusion contour covers a part of high \tanb at masses near 200 GeV and is expected to proceed to higher masses and lower \tanb values. However, the higher the charged Higgs mass, the lower the cross section. This fact may slow down the progress to heavy charged Higgs area near 0.5 TeV. In this region, the analysis presented in this work is expected to help other search analyses.

\section{Conclusions}
The $s$-channel heavy charged Higgs production was studied in the mass range $400 < m_{H^{\pm}} < 1000$ GeV, which is currently outside the reach of LHC. Using the fact that heavy charged Higgs with a mass in the range given above, produces boosted top quarks in its decay to $t\bar{b}$ pair, the top tagging technique was used to benefit from the collinearity of jets in the top quark hadronic decay. For other possible jets in the event, ususal jet reconstruction with standard cone size of 0.4 was used independently. Based on kinematic differences of signal and background, selection cuts were applied to select signal events. The increasing top tagging efficiency at higher charged Higgs masses, partially compensates the decrease in signal production cross section. This fact helps observation of charged Higgs at TeV scale. Results show that using this production process, at a large part of parameter space, an observable signal can be extracted from SM background and the mass measurement is possible in the range 0.5 to 1 TeV.
\section*{Acknowledgements}
We would like to thank Dr. Mogharrab for his continuous care and maintenance of the computing cluster at Science Department of Shiraz University, without whom and his efforts, this work would not reach its final stage.
\clearpage

\providecommand{\href}[2]{#2}\begingroup\raggedright\endgroup

\end{document}